\documentclass[aps,prd,groupedaddress,nofootinbib,letterpaper]{revtex4}

\usepackage{amsmath}
\usepackage{amsfonts}
\usepackage{amssymb}
\usepackage{amsthm}

\usepackage{braket,mathtools}
\usepackage{mathrsfs}
\usepackage{stmaryrd}

\usepackage{hyperref}
\usepackage{color}

\newcommand{\call}{{\cal L}}
\newcommand{\calo}{{\cal O}}

\newcommand{\beq}{\begin{equation}}
\newcommand{\eeq}{\end{equation}}
\newcommand{\bea}{\begin{eqnarray}}
\newcommand{\eea}{\end{eqnarray}}
\newcommand{\hf}{\frac{1}{2}}

\newcommand{\rdp}{{r^{\prime\prime}}}

\newcommand{\ehat}{{\hat e}}

\begin{document}
\begin{titlepage}

\title{Gauge-invariant observables in gravity and electromagnetism:\\  black hole backgrounds and null dressings}

\author{Steven B. Giddings}
\email{giddings@ucsb.edu}
\affiliation{Department of Physics, University of California, Santa Barbara, CA 93106\footnote{Current address}}
\affiliation{CERN, Theory Department,
1 Esplande des Particules Geneva 23, CH-1211, Switzerland}

\author{Sean Weinberg}
\email{sean.weinberg@qcware.com }
\affiliation{Department of Physics, University of California, Santa Barbara, CA 93106}

\begin{abstract}
We address  questions regarding construction and implications of gauge-invariant ``dressed" observables in nontrivial background geometries such as that of a black hole.  Formally, such observables can be constructed, {\it e.g.} by locating points with geodesics launched from infinity.  However, practical complications arise in non-trivial geometries, and in particular for observables behind black hole horizons.  Greater simplicity can be achieved by considering null constructions where the dressing lies along a null geodesic, or  null surface such as a cone.  We first investigate basic properties of these null dressings in the simpler context of electromagnetism.  Since null constructions provide simple dressings for gauge-invariant observables inside black holes, they also allow us to investigate the question of compatibility of observables inside and outside black holes, and in particular the idea of black hole complementarity.  While such observables in general have non-vanishing state-dependent commutators, the failure to commute does not appear particularly enhanced by the presence of the horizon.  

\end{abstract}

\maketitle

\end{titlepage}

\section{Introduction, motivation, and summary}

If gravity can be described in a quantum-mechanical framework, this in particular is expected to mean that it has a linear space of states (``Hilbert space"), and other basic mathematical structure of a quantum-mechanical system  including an algebra of observables.\footnote{For some more discussion, see \cite{UQM}.}  The question of how to correctly describe and interpret the quantum observables of the theory is a centrally important one in its study.

In fact, a fundamental question, from a ``quantum-first" perspective\cite{QFG,QGQFA}, is what additional mathematical structure on the Hilbert space is specifically needed to describe a quantum theory of gravity, and the observables are expected to provide important clues.  This can be appreciated by considering the analogous question of what mathematical structure is present on the Hilbert space underlying local quantum field theory (LQFT).  In LQFT, 
 the algebra of observables has the additional structure of a network of subalgebras, and this  directly corresponds to the topological structure of the underlying spacetime manifold:  subalgebras of observables localized to spacelike separated regions commute.\footnote{See, {\it e.g.}, \cite{Haag}.}  This is  key in formulating a notion of localization of information, and a corresponding definition of quantum subsystems.

While we expect this manifold structure to emerge in the limit of vanishing Newton's constant $G$ (which can be thought of as a restriction to considering states with weak gravitational fields), we find significant modifications for nonvanishing $G$.  This can begin to be seen by working perturbatively about the $G=0$ LQFT limit.  Here, one immediately finds that there are no longer local observables\cite{Torre:1993fq,DoGi2}, raising the important question of how to think about localization of information, or separability\cite{QFG} -- a subtle topic in a theory of gravity.

So far, we 
have only limited knowledge of observables in gravity, mainly from working in the small-$G$ limit. 
One can think of defining gauge invariant observables in gravity via two different approaches.\footnote{Further reflection reveals that these are more similar than  is described here.}  The first is to work relationally, and locate an observable with respect to other features of the state in question, {\it e.g.} described in terms of features of the quantum fields.  A second approach is to start with a field observable of LQFT, and then ``dress" it with a gravitational field, to make it gauge invariant.   This paper focuses on the latter approach.  

One specifically finds that gravitationally-dressed observables generically no longer commute at spacelike separations\cite{SGalg,DoGi1}, undermining the LQFT definition of locality.  Nonetheless, at least perturbatively, it appears to be possible to develop a notion of localization of information in terms of a mathematical structure on the Hilbert space called a gravitational splitting\cite{DoGi4,SBGPGS}, related to the construction of dressed observables.

Dressed observables also play important roles in related proposals for formulating quantum gravity.  A prominent example is that of AdS/CFT, where it has been argued that the correspondence between bulk and boundary observables (the ``holographic dictionary") relies on solving the gravitational constraint equations\cite{Maroholo1,Maroholo2}.  As we will review, dressed observables provide such solutions, since they are constructed to be gauge invariant, and hence commute with the constraints.

Previous concrete study of dressed observables has been limited to perturbative study about flat backgrounds\cite{DoGi1}\cite{DoGi2,DoGi3} and about AdS\cite{GiKi}.  While we of course would like to understand the story in states corresponding to general spacetimes, particularly interesting questions arise in the presence of a black hole (BH).  Here, there have been many discussions about a possibly radically different role for observables inside the black hole, epitomized by statements of BH complementarity\cite{STU,BankComp,BaFiComp}, the notion that somehow observables inside a BH are fundamental incompatible with observables outside.  More generally, there is the important question of whether a BH can be thought of, at least approximately, as a quantum subsystem.  Suggested counterarguments include those based on discussions of soft hair\cite{Hawk,HPS1,HPS2,HHPS}, where it is for example argued that information is not localized inside the BH, but is also accessible outside via the soft hair.

These questions can begin to be probed by constructing gauge-invariant observables associated to matter inside a BH.  For example, the latter question of localization of information and subsystems is addressed perturbatively in a flat background by constructing dressings that are insensitive to the state inside a certain region\cite{DoGi4,SBGPGS}.  One expects that these arguments should hold equally well if that region is inside a BH, and construction of corresponding observables and dressings is a first step towards explicitly demonstrating that.  Likewise, if there is a fundamental obstruction to the compatibility of observables inside and outside a BH, that should be seen in a mathematical description of the Hilbert space and observables.  Indeed, it was suggested in \cite{BankComp,BaFiComp} that such behavior would be seen in solutions to the Wheeler-DeWitt equation, and as we have stated, the dressed observables provide perturbative solutions.  So, a test of this question in the context of perturbative gravity can be formulated in terms of such observables.  And of course explicit construction of such interior observables also addresses concerns that have been expressed over whether they exist\cite{HaOo}.  

This paper will provide initial study of the question of defining such dressed observables in a BH background, and in particular its interior.  After reviewing general aspects of dressed observables, we discuss approaches to those inside BHs, based on using geodesics to localize points, in section III.  One can in principle consider spacelike or timelike geodesics to localize an operator inside a BH, but the resulting constructions are rather complicated.  However, significant simplification results from instead localizing along null geodesics.  We give an explicit perturbative construction of such operators, demonstrating that leading-order perturbatively-dressed operators corresponding to excitations in a BH interior exist.  Since this is a relatively new form of dressing, we investigate basic properties of null dressings in the simpler example of QED in section IV.  In particular, studying the commutators of such dressed operators is an important warm-up problem to studying commutators of null-dressed gravitational operators in BH backgrounds.  Important features of the latter commutators are then inferred, {\it e.g.} probing complementarity, in section V.

\section{Gravitational dressings: generalities}

We are interested in the problem of defining gauge-invariant observables for quantum gravity.  In general this problem entails a number of challenges.  To avoid some of them, this paper will investigate this question perturbatively, expanding the metric about a fixed background.  This approach is expected to yield important clues about the more general problem.

Consider matter coupled to Einstein gravity, with action
\beq
S=\int d^4x \sqrt{|g|} \left(\frac{2}{\kappa^2} R + \call_{gf} + \call_m\right)\ 
\eeq
where $\kappa^2=32\pi G$ is the gravitational coupling, $R$ the curvature scalar, $\call_{gf}$ a possible gauge-fixing action, and $\call_m$ a matter lagrangian.
For simplicity, we will primarily focus on the case of a scalar field,
\beq
\call_m = -\hf\left[ (\nabla \phi)^2 + m^2\phi^2\right]\ .
\eeq
The metric can be expanded about a given background $g_{\mu\nu}$ as 
\beq
{\tilde g}_{\mu\nu} = g_{\mu\nu} + \kappa h_{\mu\nu}\ ;
\eeq
for simplicity of notation, from now on we denote the full, fluctuating, metric as ${\tilde g}_{\mu\nu}$, and the fluctuation $h_{\mu\nu}$ can be perturbatively expanded in $\kappa$.  Under  infinitesimal diffeomorphisms with parameter $\kappa \xi^\mu$, the scalar field transforms as 
\beq\label{phixm}
\delta_{\kappa \xi} \phi = -\kappa \xi^\mu \partial_\mu\phi\ ,
\eeq
and the metric fluctuation transforms as
\beq\label{metxm}
\delta_{\kappa \xi} h_{\mu\nu} = -\nabla_\mu\xi_\nu - \nabla_\nu\xi_\mu\ .
\eeq

While in the quantum field theory limit, $G\rightarrow0$, $\phi(x)$ is a good quantum observable, it is no longer gauge invariant for $G\neq0$.  The problem is thus to find expressions that define gauge-invariant observables .  Such expressions were found in \cite{DoGi1}\cite{DoGi3}\cite{GiKi}\cite{QGQFA}\cite{DoGi4},\footnote{See also the earlier work \cite{Heem,KaLiGrav}.} by ``dressing" the $G=0$ observable $\phi(x)$ with a ``gravitational dressing" that creates a non-trivial gravitational field.  These expressions were extended to dressings of more general operators in \cite{DoGi3,DoGi4}.

The basic procedure is to find a functional $V^\mu[\tilde g_{\mu\nu},x]$ of the metric, describing the dressing, such that 
\beq\label{Phidef}
\Phi(x)=\phi(x + V(x))
\eeq
is diffeomorphism invariant.  We will work to leading order in the $\kappa$ expansion, and to this order, we see from \eqref{phixm} that
diffeomorphism invariance of $\Phi(x)$ follows if $V^\mu$ transforms by the key dressing relation
\beq\label{keyreln}
\delta_{\kappa \xi} V^\mu(x) = \kappa \xi^\mu(x)\ .  
\eeq

\subsection{Flat backgrounds}

We begin by reviewing dressings in a flat background, $g_{\mu\nu}=\eta_{\mu\nu}$.
Various dressings can be found expanding about this background to leading order in $\kappa$.  One of the simplest is the {\it gravitational line} construction\cite{DoGi1}\cite{QGQFA}.  Given $x$, choose a curve $\Gamma$ connecting $x$ to infinity.  Then,
\beq\label{gline}
V_\mu^\Gamma(x)= {\kappa\over 2} \int_\Gamma dx^{\prime\nu} \left\{ h_{\mu\nu}(x') + \int_{\Gamma,x'} dx^{\prime\prime\lambda}\left[\partial_\mu h_{\nu\lambda}(x'') - \partial_\nu h_{\mu\lambda}(x'')\right]\right\}
\eeq
defines this line dressing, where the second integral is along $\Gamma$ from $x'$ to $\infty$.  Under the transformation \eqref{metxm}, $V^\mu_\Gamma$ is easily checked to satisfy the key dressing relation \eqref{keyreln}.

A more symmetric, Coulomb-like, dressing may be found\cite{DoGi1} by taking $\Gamma$ to be a straight line from $x$ to infinity, and then averaging over all such lines.  This dressing creates a metric configuration corresponding to the linearized Schwarzschild solution.  The resulting expression can be extended to describe a dressing creating a more general linearized metric\cite{SBGPGS}.  Specifically, suppose that we have an $\check h_{ij}$ satisfying
\beq\label{dderiv}
\partial_i\partial_j{\check h}^{ij}_{\vec x}(\vec x')= -\delta^3(\vec x'-\vec x)\ ,
\eeq
and that 
\beq\label{Chdef}
\gamma _{\mu,ij}=\frac{\kappa}{2} \left(\partial_i h_{\mu j} +\partial_j h_{\mu i}- \partial_\mu h_{ij}\right)
\eeq
is the linearized Christoffel symbol.  Then
\beq\label{gendress}
V_\mu(x)=\int d^3x' {\check h}^{ij}_{\vec x}(\vec x') \gamma _{\mu,ij}(x')\ ,
\eeq
where the integral is over the slice $x^{\prime0}=x^0$, also satisfies the key dressing relation  \eqref{keyreln}.

The operator \eqref{gendress} creates a field configuration with initial data $h_{ij}=\check h_{ij}$.  The difference between two distinct such field configurations, satisfying \eqref{dderiv}, is a pure radiation (sourceless) field.  So, for example, the gravitational line dressing \eqref{gline}, which may also be put in this form, creates a field configuration that, in the case of a fixed static source, is expected to emit radiation to infinity and asymptote to the symmetrical Coulomb-like solution in the future\cite{DoGi1}, in parallel with the decay of the analogous electromagnetic Faraday line\cite{Shab,PFS,HaJo,SGantip}.  Thus, in general, different dressings correspond to different radiation fields superposed on a given dressing, and for example have different soft charges\cite{SBGPGS}.

The dressing $V^\mu$ can also be used to dress a more general operator 
$A$ in the $G=0$ field theory.   The dressed operator \eqref{Phidef} generalizes to \cite{DoGi4}
\beq
{\hat A} = e^{i\int d^3x\, V^\mu(x)\, T_{0\mu}(x)}\ A\ e^{-i\int d^3x\, V^\mu(x)\, T_{0\mu}(x)} \  + {\cal O}(\kappa^2)\ .
\eeq
An alternate way to check gauge invariance is to show that these dressed operators commute with the constraint operators,
\beq\label{constraints}
C_\mu(x)=G_{0\mu}(x)-8\pi G\,T_{0\mu}(x)\ ,
\eeq
which generate gauge transformations and classically vanish.  One can similarly dress a state $|\psi\rangle$ of the $G=0$ theory,
\beq
|\psi\rangle \rightarrow |\hat\psi\rangle = e^{i\int d^3x\, V^\mu(x)\, T_{0\mu}(x)} |\psi\rangle\  + {\cal O}(\kappa^2)\ .
\eeq
These states are weakly annihilated by the constraints\cite{DoGi4} \eqref{constraints}.

\subsection{General backgrounds}

An important question is how to construct dressings when perturbing about a more general background $g_{\mu\nu}$.  

A  na\"\i ve extension of the line dressing \eqref{gline} would be to, {\it e.g.}, replace the ordinary derivatives by covariant derivatives.  However, in a general background this does not yield a dressing transforming by the key dressing relation \eqref{keyreln} under diffeomorphisms.

Gravitational line dressings can be constructed in a more implicit fashion.  Suppose that the spacetime has a fixed asymptotic ``platform," on which we take the diffeomorphisms $\xi^\mu(x)$ to vanish.  For flat space, one may think of such a platform as sitting at a large distance\cite{DoGi1}; in AdS, the platform can be taken to be spatial infinity\cite{GiKi}.  Then, for a given point on the platform, one may choose a given direction, and launch a geodesic in that direction.   Points can be specified in a gauge-invariant way in terms of the boundary point, the direction, and the distance along the geodesic.  This construction may also be generalized by {\it e.g}, specifying an acceleration for the trajectory.

In the case of flat space, the geodesic construction yields the gravitational line \eqref{gline}, with $\Gamma$ a straight line\cite{DoGi1}.  A similar construction can be given in AdS\cite{GiKi}, and can be solved for an explicit expression that generalizes the gravitational line \eqref{gline}, involving geodesics in AdS.  In a more general spacetime, one would similarly like explicit constructions, but it is less clear how to write down the corresponding expressions.

One might also try to generalize the expression \eqref{gendress}, which creates a dressing that is a perturbation of a more general background metric.  However, once again, in a general background, writing an expression that satisfies the key relation \eqref{keyreln} becomes more nontrivial, due to nonzero background curvature.

\section{Dressings in black hole backgrounds}

In seeking generalizations of the dressings \eqref{gline} or \eqref{gendress} to more general spacetimes, we will try to take a modest step beyond Minkowski and AdS spacetimes and consider black hole spacetimes.  This also introduces new questions, associated with the nontrivial causal structure of BH spacetimes, and in particular the existence of horizons.

\subsection{Spacelike dressings}

In Minkowski space, the simplest way to construct a diffeomorphism-invariant observable at leading order is, as described above, to attach a nonlocal spacelike dressing to a local quantum field.  
This procedure is also possible in other backgrounds.  For example, consider a static black hole in Schwarzschild coordinates $(t, r, \theta^A)$, with $\theta^A=\theta,\phi$.  If we restrict only to gravitational
line dressings which are radially directed at constant time, then it is not very difficult to find $V_\mu$ satisfying the key relation \eqref{keyreln}.   The resulting operators, $\Phi_{\rm static}(x)$,  are well-defined, but only outside of the event horizon.  

These observables can be specified by giving explicit expressions for the corresponding dressing operators $V_\mu(x)$.
Let $f(r) = 1 - 2 GM / r$.  Then, $\Phi_{\rm static}$ has the dressing
\bea
\label{static_dress}
V_t(x) &=& \kappa f(r) \int_r^\infty dr' \frac{1}{f(r')} \left[ h_{tr}(t,r',\theta^A) + \frac{1}{2 \sqrt{f(r')}} \int_{r'}^\infty d\rdp \sqrt{f(\rdp)}  \partial_t h_{rr}(t,\rdp,\theta^A) \right]\cr
V_r(x)&=& \frac{\kappa}{2 \sqrt{f(r)}} \int_r^\infty dr'  \sqrt{f(r')} \:h_{rr}(t,r',\theta^A)\cr
V_A(x)&=& \kappa \int_r^\infty dr'\left(\frac{r}{r'}\right)^2\left[ h_{A r}(t,r', \theta^B)+ \frac{1}{2 \sqrt{f(r')}} \int_{r'}^\infty d\rdp \sqrt{f(\rdp)} \partial_A h_{rr}(t,\rdp,\theta^B)\right]
\eea

This static dressing is manifestly problematic at the black hole horizon.  On the other hand,
the construction appears to also work inside of the future (or past) horizon of a black hole, but in this case the gravitational line is timelike rather than spacelike.  Moreover, in this situation, the dressing terminates at the black hole singularity rather than at infinity, and it is thus not  clear in what sense the operator should be considered gauge-invariant.  

While this construction has the advantage of relative simplicity, we will largely avoid it below because of the major flaw that it does not directly address questions about the relationships between observables inside and outside of the horizon. There is a discontinuity in the static construction between operators in the interior and exterior of a black hole, and, moreover, there is no way to talk about an operator that originates at a point on the event horizon itself.
\begin{figure}
\begin{centering}
\includegraphics[scale=0.7]{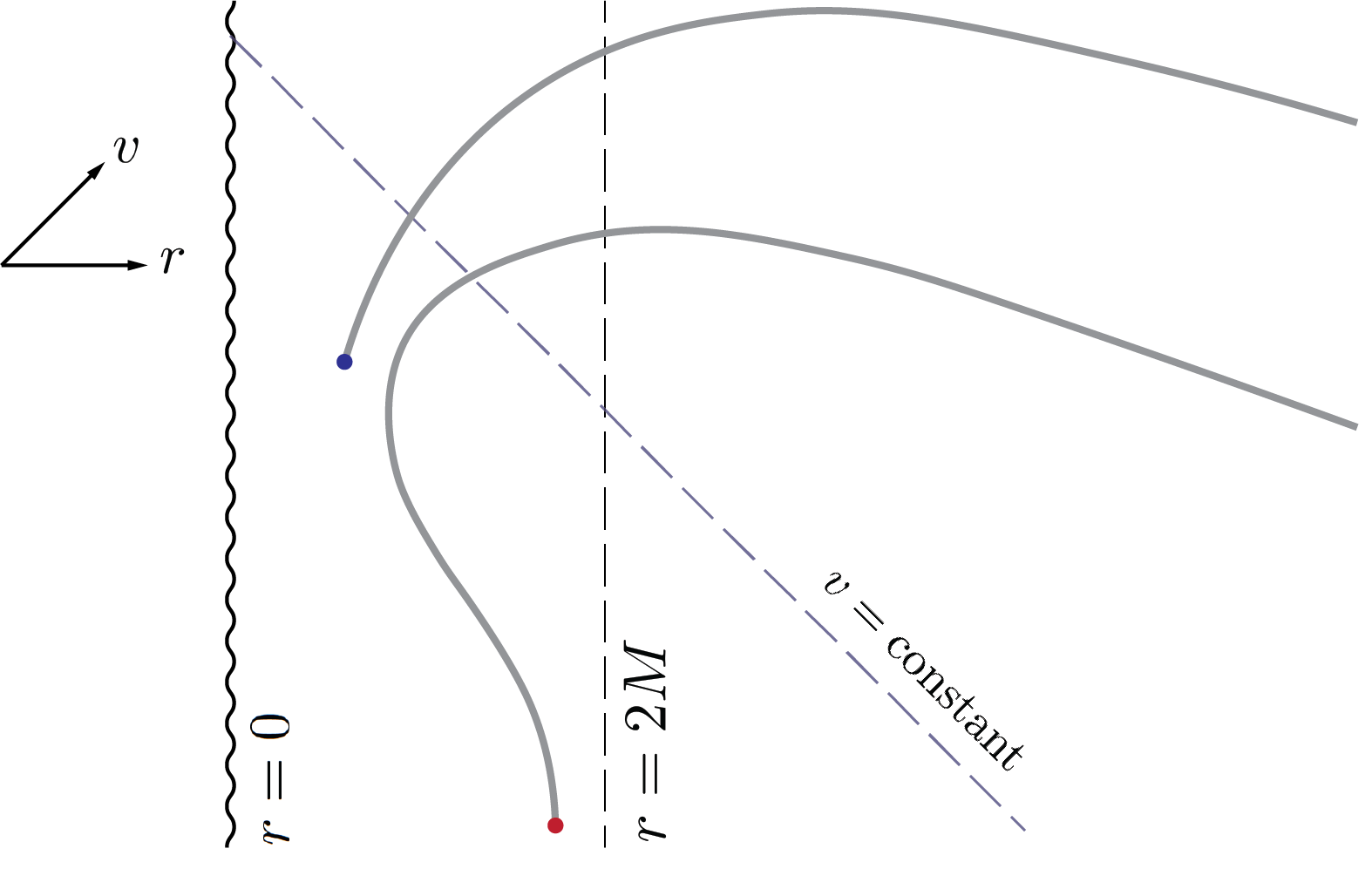}\caption{\label{fig:ef}As shown in ingoing Eddington-Finkelstein coordinates, we see that non-static spacelike dressings  emerge from an initial value of $v$, propagate to a minimal positive value of $r$, and finally emerge from the black hole.  The awkwardness of the dressings contributes to our choice below to use null dressings instead for detailed calculations.}
\par\end{centering}
\end{figure}

\begin{figure}
\begin{centering}
\includegraphics[scale=0.7]{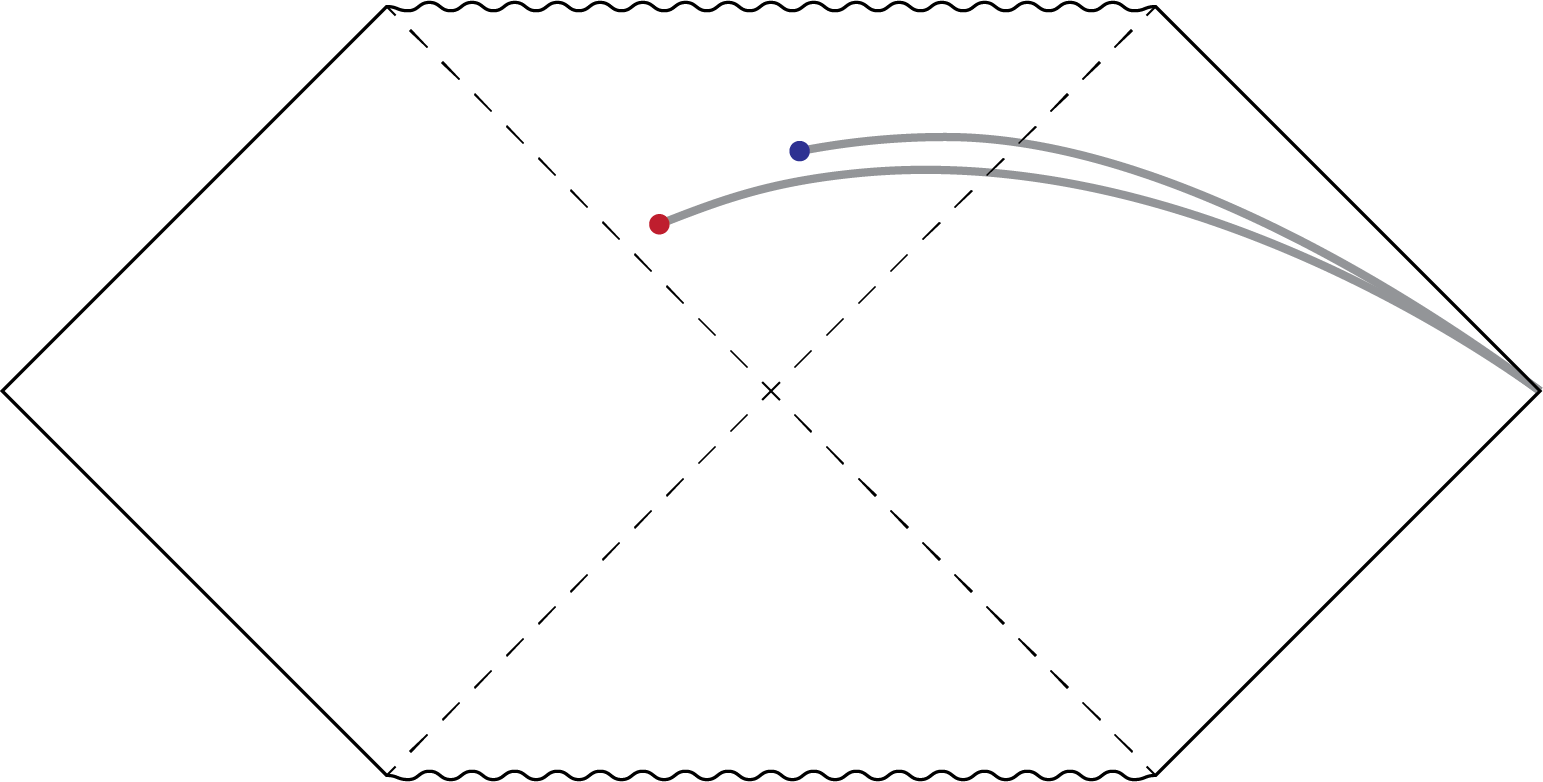}\caption{\label{fig:penrose} The same pair of geodesics that are shown in figure \ref{fig:ef} are shown here in a maximally extended geometry.}
\par\end{centering}
\end{figure}
This appears not to be a fundamental problem, and a solution is to give up on dressing operators only on static radial curves.  By considering spacelike geodesics launched with non-zero rapidity from infinity, one can reach points inside the horizon.  These geodesics then allow construction of gauge-invariant operators interior to the black hole.  Such operators are schematically illustrated in figs.~\ref{fig:ef} and \ref{fig:penrose}.
 These operators are  smooth functions of their originating point and are not sensitive to the event horizon in any dramatic way as long as a reasonable family of spacelike geodesics are chosen.

Given observables constructed in this way, we are in a position to explore their properties, such as commutators of diffeomorphism-invariant observables inside and outside a black hole.  These investigations put further constraints on discussions of properties of black holes, their quantum observables, and properties of quantum information in the black hole context.

Unfortunately, proceeding in this direction leads to difficult computational issues.  The explicit form of operators with gravitational dressing defined via spacelike geodesics that do not respect the spacetime's timelike Killing vector field is difficult to write down and even more cumbersome to extract interesting results from.  A primary challenge of working with such operators is related to the structure of spacelike geodesics in a black hole background.  

Figures \ref{fig:ef} and \ref{fig:penrose} illustrate examples of local operators that have been gravitationally dressed with spacelike geodesics that are not static.  These dressings do not hit the black hole singularity.  The geodesics can also be extended into the second (unphysical) exterior region of the Schwarzschild spacetime. 

While ingoing Eddington-Finkelstein coordinates are often  useful for giving a non-distorted view of the spacetime, localizing points with these geodesics looks particularly awkward in such a description; for example, a radially ingoing geodesic starting outside of a black hole propagates to $v=-\infty$ (see figure \ref{fig:ef}). This awkward construction, combined with the lack of a clean way to explicitly parameterize the geodesics as functions of proper length (which is a necessary step when constructing gauge-invariant operators in perturbative quantum gravity) motivates exploring different approaches. 

\subsection{Timelike dressing}

The challenges of the use of spatial geodesics for constructing dressings  in the full spacetime  suggests exploration of other approaches.  One possibility that leads to a natural way to probe the interior of horizons is to instead consider timelike geodesics.  For example, one may choose a family of freely-falling observers from infinity, which ultimately cross the horizon.  These define a family of timelike geodesics crossing the horizon, and specifically points inside the horizon may be localized by specifying an observer (for example defined in terms of a frame at infinity) and the proper distance along that observer's trajectory. 

An example of such a construction is that of the Painlev\'e-Gullstrand coordinates.\footnote{For a review and further references, see, {\it e.g.}, \cite{MaPo}.}  In this case, the family is of observers with vanishing velocity at infinity; such observers then define spatially flat time slices, and the Schwarzschild metric becomes
\beq
ds^2=-\left(1-\frac{2GM}{r} \right) dT^2 + 2 \sqrt{\frac{2GM}{r} }dT dr + dr^2 + r^2 d\Omega^2\ .
\eeq
Thus, one prescription is to use the dressing associated with such a  choice of coordinates or gauge, generalized to a perturbed spacetime.  

The resulting construction also appears somewhat complicated; we would like to have a construction of gauge invariant observables which permits description of both sides of the event horizon, and {\it also} has the computational simplicity of the static construction.  We will leave further exploration of such timelike dressings for future work, but will instead, in the next section, work with the even simpler case of  quantum fields that are dressed with null geodesics rather than spacelike or timelike geodesics.  These dressings are very easy to write down (see equation \eqref{nulldress}) and can be used to construct a fairly rich collection of observables that reveal features of the algebras of observables for black hole backgrounds.

\subsection{Null dressing}

As the preceding sections have outlined, a conceptually simple way to specify a diffeomorphism-invariant operator is via a geodesic construction, locating the operator in terms of distance along a particular geodesic from infinity.  These yield a gravitational line dressing.  However, as the black hole case begins to show, in practice such constructions can be rather complicated, due to non-trivial behavior of geodesics.  

In certain cases, like that of a static black hole, we expect null geodesics to have  simplifying aspects, related to the utility of null coordinates.  For that reason, we turn to investigation of dressings formulated via null geodesics.

In particular, consider the Schwarzschild solution in ingoing Eddington-Finkelstein coordinates $x=(v,r,\theta^A)$,
\beq\label{SchEF}
ds^2=-\left(1-\frac{2GM}{r} \right) dv^2 + 2 dv\,dr + r^2 d\Omega^2\ 
\eeq
and an operator $\phi(v,r,\theta^A)$.  Then the corresponding null line-dressed operator is of the form \eqref{Phidef}, with the null line dressing given by
\bea\label{nulldress}
V_v(x) &=& \kappa \int_r^\infty dr'\left\{ h_{rv}(v,r',\theta^A) + \hf \int_{r'}^\infty d\rdp \left[ \partial_v h_{rr}(v,\rdp,\theta^A) + \frac{2M}{r^{\prime 2}} h_{rr}(v,\rdp,\theta^A)\right]\right\}\cr
V_r(x)&=&\frac{\kappa}{2}\int_r^\infty dr' h_{rr}(v,r',\theta^A)\cr
V_A(x)&=& \kappa \int_r^\infty dr'\left(\frac{r}{r'}\right)^2\left[h_{rA}(v,r'\theta^B)+\hf\int_{r'}^\infty d\rdp \partial_A h_{rr}(v,\rdp,\theta^B)\right]\ .
\eea
This dressing can be checked to satisfy the key dressing relation \eqref{keyreln}.  The expressions also make sense whether or not the operator lies inside the BH horizon at $r=2M$.  

The dressings in \eqref{nulldress}, and others like them, can be obtained in a variety of ways without attempting to rely on a generalization of the general line dressing expression, eq.~\eqref{gline}.  One approach is to
consider that the operator $\Phi(x) = \phi(x + V(x))$ is gauge invariant if $x + V(x)$ refers  to the same spacetime point before and after a coordinate transformation.   From this perspective, $V$ is an operator 
that always implements the appropriate diffeomorphism.  As with other geodesics, this is accomplished
in a  direct way by, for example, labeling spacetime points by null affine distance (with a fixed normalization rule) from a fixed ``platform'' at infinity.  If such coordinates are used, we will always have $g_{\lambda\alpha}$ constant
where $\lambda$ is affine parameter and $\alpha$ refers to coordinate system on the platform transverse to the coordinate $\lambda$.  In our expressions above, $r$ plays the roll of $\lambda$.  The  value of $g_{r \alpha}$
 should be fixed even when we allow the metric to fluctuate, so under a gauge transformation \eqref{metxm}, we  require that
\beq
\label{axial}
 h_{\alpha r} -\nabla_\alpha\xi_r - \nabla_r \xi_\alpha\  = 0.
\eeq
This first-order differential equation can be solved by integration to obtain a relationship between $\xi$ and $h$.  $V$ is  found by letting it play the role of $\xi$ in \eqref{axial} and by interpreting $h$ as the metric fluctuation operator.

The null dressing \eqref{nulldress} provides an example of the general class of null dressings based on null geodesics.  It is important to understand the properties of these dressings.  Particular questions include the behavior of the dressing fields created by these dressings, and their relation to those of more ``conventional" dressings based on spacelike geodesics, as well as  the form of the commutators of such dressings, either with other null dressings, or with other types of dressing.

Notice that such null dressings, which will be studied further below, provide an example of operators that are not localized to spacelike time slices.  This suggests a possible illustration of circumvention of some of the issues raised in \cite{Jaff}, regarding non equal-time dressings, but more detailed discussion of this question will be left for future work.  We  also have noted that these null-dressed operators can be defined for fields well inside the horizon, and as we will discuss, they appear to have reasonable properties there.  They therefore appear to  show  explicitly how to circumvent statements made in \cite{HaOo} (see sec. 4.2), that dressed operators shouldn't extend deep within a black hole.

Since the expressions \eqref{nulldress} are still somewhat complicated, and expressions such as commutators become even more complicated in the nontrivial metric 
\eqref{SchEF}, we will  first approach the question of understanding properties of null dressings with simpler cases, beginning with that of electromagnetic dressings in flat space.

\section{Dressings in QED}\label{QEDdress}

A simpler example than gravity for many properties of dressings is that of QED.  Consider electromagnetism coupled to a scalar, with lagrangian
\beq
\call_{\rm EM} = -\frac{1}{4} F^{\mu\nu}F_{\mu\nu}-|D_\mu\phi|^2 - m^2|\phi|^2\ .
\eeq
Gauge transformations act as
\beq\label{gaugexm}
\phi(x)\rightarrow e^{-i q \Lambda(x)}\phi(x)\quad ,\quad A\rightarrow A-d\Lambda\ ,
\eeq
and the field strength and gauge-covariant derivative are
\beq
F_{\mu\nu}=\partial_\mu A_\nu- \partial_\nu A_\mu\quad ,\quad D_\mu\phi = \partial_\mu\phi -iq A_\mu\phi\ .
\eeq

In order to quantize, one needs a gauge fixing prescription.  One particular choice is to add a Feynman gauge-fixing (or breaking) term,
\beq
\call_{\rm gf}=-\frac{1}{2\alpha} (\partial_\mu A^\mu)^2\ .
\eeq
Then, the canonical momenta are
\beq
\pi^i=-F^{0i}=-E^i\quad ,\quad \pi^0=\frac{1}{\alpha}\partial_\mu A^\mu\ ,
\eeq
and the canonical commutators are
\beq\label{cancom}
[\pi^\mu(x), A_\nu(x')]_{\big|_{t=t'}}= -i \delta_\nu^\mu \delta^3(\vec x -\vec x')\quad ,\quad [A_\mu(x), A_\nu(x')]_{\big|_{t=t'}}=[\pi^\mu(x), \pi^\nu(x')]_{\big|_{t=t'}}=0\ .
\eeq

\subsection{General dressings}

The gauge-variant operator $\phi(x)$, which can be thought of as creating a charged particle at $x$, can be dressed to give an operator
\beq\label{PhiEM}
\Phi(x) = \phi(x) e^{iqV(x)}
\eeq
where the electromagnetic dressing $V(x)$ is a functional of $A_\mu$.  This dressed operator will be gauge invariant if under \eqref{gaugexm} (compare \eqref{keyreln})
\beq
V(x)\rightarrow V(x) + \Lambda(x)\ .
\eeq
The dressed operator \eqref{PhiEM} can be thought of as creating the particle, together with its EM field, in some configuration determined by the particular choice of dressing $V(x)$.  There are infinitely many such possible dressings.\footnote{For previous discussions of EM dressings, see \cite{Dirac1955,Bucholz1982,Steinmann1983,Steinmann2004}\cite{SGantip}.}  A general class is of the form\cite{DoGi1}
\beq
V(x)=  \int d^4x' f^\mu(x,x') A_\mu(x')\ ,
\eeq
where $f^\mu(x,x')$ is a function satisfying
\beq\label{genEMdress}
\partial'_\mu f^\mu(x,x') = \delta^4(x-x')\ .
\eeq

One specific dressing is the Faraday line dressing given by Dirac\cite{Dirac1955} (compare \eqref{gline}),
\beq\label{Faraday}
V=\int_\Gamma A
\eeq
where $\Gamma$ is a curve from $x$ to infinity, at constant time $t$.  A more general example is of the form (compare \eqref{gendress})
\beq\label{genEM}
V(t,\vec x)=\int d^3 x' {\check E}^i_{\vec x}(\vec x') A_i(t,\vec x')\ ,
\eeq
where $\check E^i_{\vec x}$ is the classical electric field of a unit charge at $\vec x$,
\beq
\nabla_i' \check E^i_{\vec x}(\vec x')= \delta^3(\vec x'-\vec x)\ .
\eeq
The special case \eqref{Faraday} corresponds to the situation where the electric field lines all lie along $\Gamma$.\footnote{A related construction given by Mandelstam\cite{Mand} is to run such a line back in time to a fiducial time, and then to infinity in the corresponding spatial slice, rather than running the line to infinity at equal time to the original operator.  These operators have different properties, such as commutators.  We will explore similar constructions below.  This kind of operator has also been used in work investigating dressing of particle states, {\it e.g.} recently in \cite{CPA}.}  

The different choices of dressings give operators that create EM field configurations that differ from one another by a homogeneous (sourceless) solution of the Maxwell equations.  That is, in the situation where two differently dressed operators \eqref{PhiEM} create particles, they also create EM fields for those particles that differ by a pure radiation field.  The differently-dressed operators also have different soft charges\cite{SBGPGS}.

To see more clearly the EM field configurations  created by the  operators with dressing \eqref{genEM}, consider the equal-time commutator of the dressed operator \eqref{PhiEM} with the field operators, using the commutation relations \eqref{cancom}:
\beq\label{Dcomms}
[A_\mu(t,\vec x'),\Phi(t,\vec x)] = 0\quad,\quad [E^i(t,\vec x'),\Phi(t,\vec x)] = -q {\check E}^i_{\vec x}(\vec x') \Phi(t,\vec x) \quad,\quad [\partial_\mu A^\mu(t,\vec x'),\Phi(t,\vec x)]=0\ .
\eeq		
Since the commutators are proportional to the original dressed operator, one may clearly identify the coefficients as the dressing fields created by the dressings.  The third relation in \eqref{Dcomms} shows that in our quantization scheme, the dressing field is in Lorenz gauge at time $t$.  This extends to general times\cite{DoGi1}, since $\partial^\mu A_\mu$ generates gauge transformations, and $\Phi(x)$ is gauge invariant.

A particularly simple dressing is found by averaging \eqref{Faraday} over all straight lines from $x$ to infinity, to give the dressing
\beq\label{Couldress}
V_D(0)=\frac{1}{4\pi}\int d\Omega dr e^i A_i(t=0, r, e^i)\ ,
\eeq
where $e^i$ is the unit radial vector.  This ``Dirac dressing" was first given in \cite{Dirac1955}.  This is also of the general form \eqref{genEM}, with an $\check E^i$ corresponding to a Coulomb field.
The corresponding gauge potential, for $x'$ spacelike separated from $x$, was found at leading order in $q$ in \cite{DoGi1},
\beq
[A_\mu(x'),\Phi_D(x)] = \tilde A_{D\mu}(x') \Phi_D(x)
\eeq
with
\beq\label{Coulpot}
\tilde A_{D\mu}(x')=\frac{q}{4\pi} (t'-t) \delta_\mu^i \frac{(\vec x' - \vec x)_i}{|\vec x' - \vec x|^3}\ .
\eeq

\subsection{Null dressings}\label{NullDr}
\label{subsec:null_dressing}

We have motivated an interest in null dressings, and there is an obvious null generalization of the Faraday line \eqref{Faraday}:  one simply takes the curve $\Gamma$ to be a null line from $x$ to infinity, {\it e.g.} along a past-directed null geodesic.  In order to better understand properties of null dressings, we would like to understand what kind of dressing field such operators create, and what kinds of commutators the null-dressed operators obey.

For simplicity, choose coordinates so that the operator \eqref{PhiEM} is located at $t=x=0$.  It is useful to introduce null coordinates,
\beq
u=t-r\quad,\quad v=t+r\ .
\eeq 
Then, a null line dressing in the $e^i$ direction take the form
\beq
V(0)=\int_0^{-\infty} du A_u(u,v=0, e^i)\ .
\eeq
In order to study properties of these dressings, it is simpler to consider an angular averaged dressing, 
\beq\label{nullav}
V_N(0)=\frac{1}{4\pi}\int d\Omega\int_0^{-\infty} du A_u(u,v=0, e^i)\ .
\eeq
Notice that this can also be put in a form similar to \eqref{genEM},
\beq\label{VNE}
V_N(0)= \int_0^{-\infty} (-du) \int r^2 d\Omega\,  {\check E}^\mu\, A_\mu(u, v=0, e^i)\ ,
\eeq
with 
\beq\label{Emudef}
{\check E}^\nu = -\frac{u^\nu}{4\pi r^2}=-u^\mu {\check F}_{\mu\nu}\ .
\eeq
Here $u^\mu$ is the vector in the $u$ direction,
 \beq\label{uvec}
u^\mu=\hf (1,-e^i)\ ,
\eeq
and $\check F_{\mu\nu}$ is the Coulomb field for unit charge,
\beq\label{Fdef}
{\check F}^{0i}= \frac{e^i}{4\pi r^2}\quad ,\quad {\check F}^{ij}=0\ .
\eeq

The dressing \eqref{nullav} is 
analogous to the Dirac dressing $V_D$ in \eqref{Couldress}, and a first question is what dressing field it produces.  By symmetry, we expect that this operator also creates a Coulomb field, but this may be in a different gauge than the dressing potential \eqref{Coulpot}.\footnote{In addition, at higher order in $q$, where matter loops and other higher-order corrections arise, there will be further differences.}  
Indeed, we will find that the dressing gauge potential $\tilde{A}_{N\mu}$, which arises from  $V_N$, differs from \eqref{Coulpot} only by a divergent gauge transformation, at leading order in $q$, so does yield a Coulomb field.
This divergence does raise questions about the status of the operator \eqref{nullav} and its gravitational analog discussed later.
 
 To see these properties in some detail, we begin with the commutator $[A_\mu(x'),\Phi_N(x)]$, where $\Phi_N(x)$ is of the general dressed form \eqref{PhiEM}, with dressing $V_N$ of \eqref{nullav}.  For $x'-x$ spacelike, we find, to leading order in $q$,
 \beq
 [A_\mu(x'),\Phi_N(x)] = \phi(x) [A_\mu(x'),e^{iqV_N(x)}] = iq [A_\mu(x'),V_N(x)]\, \Phi_N(x) + \calo(q^2)\ .
 \eeq
 Thus, we identify the leading order dressing potential as
 \beq
 \tilde{A}_{N\mu}(x')= iq [A_\mu(x'),V_N(x)]\ .
 \eeq
 This may be evaluated using the unequal-time leading-order commutator
 \beq
 [A_\mu(x), A_\nu(x')] = i \eta_{\mu \nu} D(x - x') + (1 - \alpha) \partial_\mu \partial_\nu E(x - x')
 \eeq
 where $x$ and $x'$ are points in Minkowski space, $D$ is the massless Pauli-Jordan function given by
 \beq
 D(x) = -\frac{1}{2 \pi}{\rm sign}(x^0)\: \delta\left(x^2\right) 
 \eeq
 and $E$ is a Green function for $\Box^2=(\partial^2)^2$:
 \beq
 E(x) = -{1 \over 8 \pi} {\rm sign}(x) \theta(-x^2)\ . 
\eeq

As previously, we choose $x=0$.  For $x'$ spacelike from the origin, we therefore have at leading order
\beq
 \tilde A_{N\mu}(x') = \frac{iq}{4\pi}\int d\Omega \: du \: u^\nu [A_\mu(x'),  A_\nu(u,v=0, e^i)]\ .
\eeq
There are two terms in the commutator, one involving $D$ and one involving $\partial_\mu \partial_\nu E$.  The latter clearly produces a pure gauge piece, and won't contribute to $\tilde F$; it can alternately be eliminated by taking $\alpha=1$, which we do at this stage.
This leaves the first term, giving
\beq
 \tilde A_{N\mu}(x') = \frac{q}{8 \pi^2}\int d\Omega \: du \: u_\mu  \delta[(x' - x)^2]\ .
\eeq

To evaluate this, note that  $(x' - x)^2 = x'^2+ u (t' + r' \cos \theta)$ where $\theta$ is the angle between ${\bold x}$ and ${\bold x}'$.  The resulting integrals have a long distance divergence, which can be regulated with a long-distance cutoff $L$.  They then give
\bea\label{nullA}
\tilde A_{N t}(x') &=& \frac{q}{16 \pi^2} \int d\Omega \int_{-\infty}^{u'} du\, \delta\left[  x'^2+ u (t' + r' \cos \theta) \right] = -\frac{q}{8 \pi r'} \log\left(u' \over -L\right)\nonumber \\
\tilde A_{N r}(x') &=& \frac{q}{16 \pi^2} \int d\Omega \int_{-\infty}^{u'} du \cos \theta \delta\left[  x'^2+ u (t' + r' \cos \theta) \right] = -\frac{q} {8 \pi r'^2}\left[ \frac{x'^2}{u'} -  t'   \log\left(u' \over -L\right)\right]\ .
\eea
This form of $\tilde A_N $ is unfamiliar and the expression diverges logarithmically as $L$, the size of the null cone, goes to infinity.  However, applying a gauge
transformation $\tilde A_N \to \tilde A_N + d \Lambda$ with
\beq
\Lambda = \frac{qu' }{8 \pi r'} \left[\log\left(u' \over -L\right) - 1 \right]
\eeq
puts the field in the form
\begin{align}
{\tilde A}_t(x') =& 0 \\
{\tilde A}_r(x') =& q\frac{ t'}{4 \pi r'^2}
\end{align}
which matches the Coulomb field expression \eqref{Coulpot}.

The divergence in the dressing gauge potential \eqref{nullA} raises the question of whether $\Phi_N$ is a well-defined member of the algebra of operators.  However, one very simple way to rectify this without losing much 
of the utility of the null dressing is to cut the null cone at a large radius and then attach to it a spacelike dressing that goes out to infinity.  This procedure renders the gauge finite and defines the operator
$\Phi_N$ in a cutoff dependent fashion.

An important lesson emphasized by this discussion, which also applies in the gravitational case, is that when an operator is rendered gauge invariant by including a dressing, various choices of dressing
can yield distinct operators.  In the present case the operators $\Phi_D$ and $\Phi_N$ both create the Coulomb field, but they do so in different gauges even though the two operators are themselves gauge-invariant!  There will also be higher-order differences between these operators since they create the Coulomb field at different times.

\subsection{Algebra and noncommutativity; state dependence}

In general it is important to understand the form of the commutators between dressed operators, since these commutators encode locality properties of the algebra of observables.  Specifically, to better understand properties of null-dressed operators, one target is to understand their commutators both in the EM and gravitational context.

Ref.~\cite{DoGi1} gave a preliminary study of some properties of such commutators, for both EM and gravity.  An example is the commutator between Dirac-dressed operators in EM,
\beq\label{Dcomm}
[\dot\Phi_D(x),\Phi_D(x')]= \frac{iq^2}{4\pi|\vec x -\vec x'|} \Phi_D(x)\Phi_D(x')\ .
\eeq

A first point, illustrated by \eqref{Dcomm}, is that in general the commutators are operators, rather than c-numbers, and so their sizes will depend on the state that they act on. One might emphasize this by calling these commutators ``state-dependent," although this simply expresses the general property that values of operators are state dependent.  But this does mean that some care is needed in discussing magnitudes of these commutators.

A second point is that the commutator is proportional to an interaction energy of particles created by the operators; we will discuss this feature further below.

\subsubsection{Shell operators}

The integrals that are needed to evaluate general commutators involving $\Phi_D$ and $\Phi_N$ are rather complicated.  For this reason, we will try to illustrate some typical properties of such commutators in cases where we preserve a significant amount of symmetry.  So, while one might in general want to study commutators of the form \eqref{Dcomm}, with general points $x$ and $x'$, we will typically translate so $x=0$.  We can then consider another operator, with undressed form
\beq\label{shell}
\phi(0, R)\ ,
\eeq
which we think of as creating a surface charge on the sphere of radius $R$ surrounding the point $ x= 0$; this operator can for example be thought of as the limit of a product $\prod_a \sigma_a$ of operators at points on the sphere.  If this operator has a total charge $q$, then it can have a dressing analogous to the Dirac dressing \eqref{Couldress}, in which case it takes the form
\beq\label{diracshell}
\Phi_D(0, R)=\phi(0, R) \exp\left\{\frac{iq}{4\pi} \int_{r>R} d\Omega dr e^i A_i(t=0, r, e^i)\right\}\ .
\eeq
This preserves the spherical symmetry about  $\vec x=0$, and also clearly trivially generalizes to a shell centered on any other point $ x\neq 0$.  

Alternately, we may dress the shell operator \eqref{shell} with a null cone dressing, analogous to \eqref{nullav}.  This takes the form
\beq\label{nullshell}
\Phi_N( 0, R)=\phi(0, R) \exp\left\{\frac{iq}{4\pi} \int_{-R}^{-\infty}du\int d\Omega \, u^\mu A_\mu(u, v=R, e^i)\right\}\ ,
\eeq
where $u^\mu$ was given in \eqref{uvec}.  We expect these dressed shell operators to illustrate important aspects of the more basic operators \eqref{PhiEM}.

\subsubsection{Commutators of shell operators}

We will  infer general aspects of the behavior of commutators by considering commutators of the shell operators \eqref{diracshell}, \eqref{nullshell},  with 
the operators $\Phi_D(x)$, $\Phi_N(x)$.
Simple first examples are the commutators $[\Phi_D(0),\Phi_D(0,R)]$, which vanishes by the canonical commutation relations \eqref{cancom}, and 
\beq
 [\dot\Phi_D(0), \Phi_D(0,R)]= \frac{iq^2}{4\pi R}\Phi_D(0,R) \Phi_D(0)
\eeq
which also follows from the equal time commutators, and is analogous to the commutator \eqref{Dcomm}.  These can also be simply generalized to commutators involving two shells, $\Phi_D(0,R_1)$ and $\Phi_D(0,R_2)$, as well as to the case with nonzero temporal separation.  

In the latter case, for small time separation, we have
\beq
[\Phi_{D}(0,R_{1}),\Phi_{D}(\delta t,R_{2})]=\left[\Phi_{D}(0,R_{1}),\phi(R_{2})\,\exp\left\{\frac{iq}{4\pi}\int_{r>R_{2}}d\Omega dr\,e^{i}A_{i}(\delta t,r,\ehat)\right\}\right]
\eeq
where we assume $\delta t \ll |R_{1}-R_{2}|$. Suppose $R_{2}>R_{1}$. Since $\delta t$ is small, the
support of the operator $\Phi_{D}(\delta t,R_{2})$ is a proper subset
of the region spacelike and outside of the sphere of radius $R_{1}$
at $t=0$, simplifying the calculation. We consider only the leading order terms in $\delta t$ and $q$, in which case the commutator becomes 
\beq
[\Phi_{D}(0,R_{1}),\Phi_{D}(\delta t,R_{2})]=\phi(R_{2})\frac{iq}{4\pi}\int_{r>R_{2}}d\Omega dr\,\,\left[\Phi_{D}(0,R_{1}),e^{i}\,A_{i}(\delta t,r,\ehat )\right]+{\cal O}(\delta t^{2},q^{2})\ .
\eeq
The integrand is evaluated using
\beq
\left[\int_{r>R_{1}}d\Omega'dr'\,e^{i\prime}A_{i}(0,\mathbf x'),e^{j}A_{j}(\delta t,\mathbf x)\right] 
  =i\int\frac{d^{3}x'}{r'^{2}}\ehat'\cdot\ehat \frac{1}{2\pi}\delta\left(-\delta t^{2}+|\mathbf{x}-\mathbf{x}'|^{2}\right)=\frac{i \delta t}{r^2}\ ,
\eeq
where the integral can be performed using methods outlined in  Appendix A.
The leading-order commutator is therefore
\bea\label{Shellcommt}
[\Phi_{D}(0,R_{1}),\Phi_{D}(\delta t,R_{2})] &=& \phi(R_{1})\phi(R_{2})\left(\frac{iq}{4\pi}\right)\int_{r>R_{2}}d\Omega dr\,\left(-q\frac{\delta t}{4\pi r^{2}}\right)+{\cal O}(\delta t^{2},q^{2})\cr
 &=& -i\frac{q^2\,\delta t}{4\pi R_{2}}\Phi(R_{1})\Phi(R_{2})+{\cal O}(\delta t^{2},q^{2})
\eea
Note that the coefficient in this expression is the integrated interaction energy, or interaction action, between the two shells.

\begin{figure}
\begin{centering}
\includegraphics[scale=0.7]{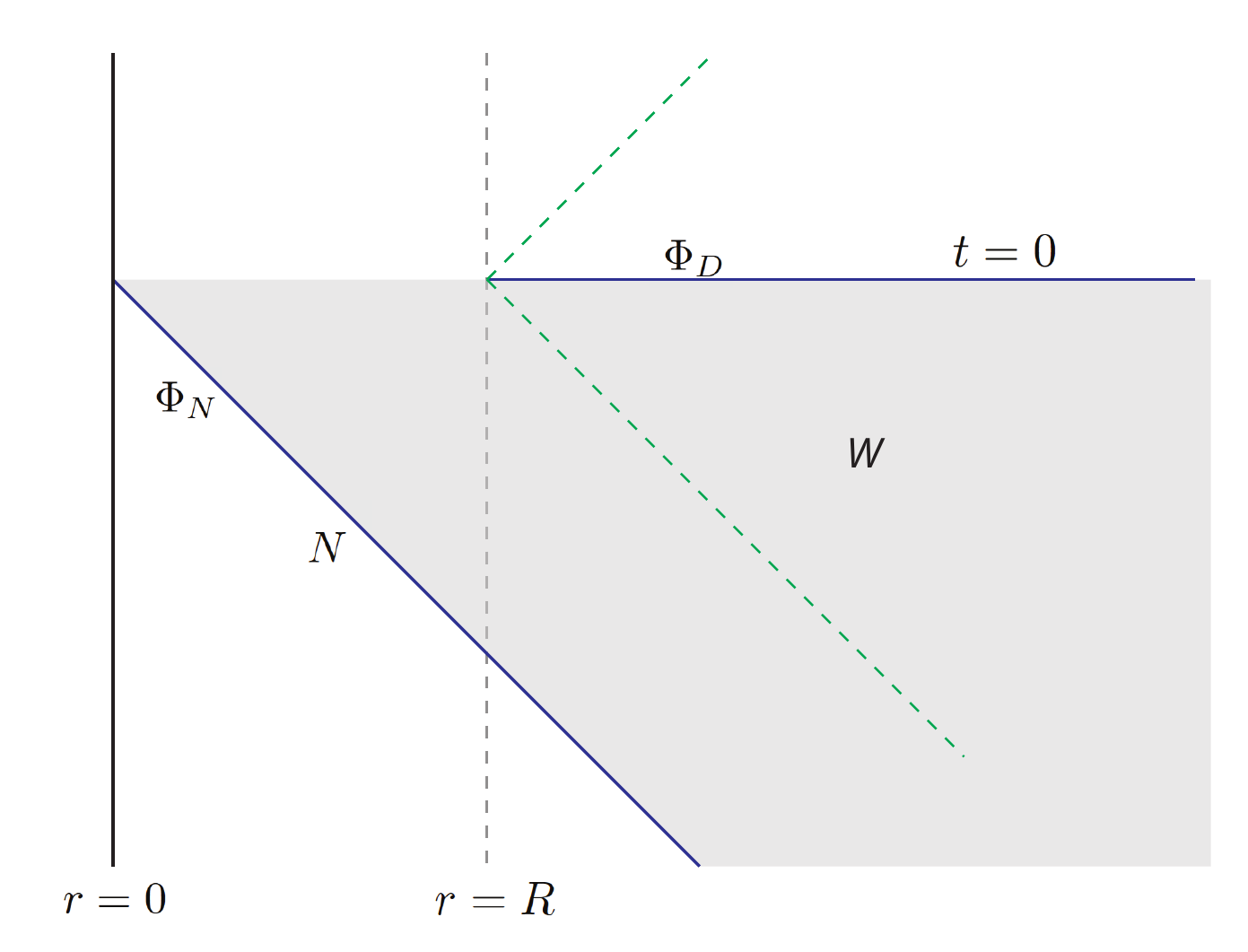}\caption{\label{fig:Wgeom}
Shown are features relevant for calculating a commutator between a null-dressed operator $\Phi_N$ at the origin, with dressing along $N$, and a Dirac-dressed shell operator $\Phi_D$ at a radius $r=R$.}
\par\end{centering}
\end{figure}

To investigate behavior of null dressings, we next turn to commutators of the form $[\Phi_N(0),\Phi_D(0,R)]$ or  $[\Phi_N(0),\Phi_N(0,R)]$, {\it e.g.} as illustrated in fig.~\ref{fig:Wgeom}.  From the general expression \eqref{PhiEM}, we find that
\beq
[\Phi_N(0),\Phi_D(0,R)]=\phi(0) [ e^{iq V_N(0)},\Phi_D(0,R)] = i \phi(0) [ qV_N(0),\Phi_D(0,R)] + \calo(q^2)
\eeq
and likewise with the replacement $D\rightarrow N$.  Using the dressing \eqref{VNE}, this becomes
\beq\label{NDcom}
[\Phi_N(0),\Phi_D(0,R)]=i \phi(0) \int_0^{-\infty} du \int r^2 d\Omega\,  u^\mu \,q{\check F}_{\mu\nu}\, [A^\nu(u, v=0, e^i),\Phi_D(0,R)]+ \calo(q^2)\ ,
\eeq
where ${\check F}^{\mu\nu}$ was given in \eqref{Fdef}.

As was discussed above, commutators such as \eqref{NDcom} are nontrivial operators.  This means that their values are state dependent, and so to quantify their size we need to choose states between  which to evaluate their matrix elements.

For example, if $|\psi\rangle$ is a state with charge $q$ -- so approximately a one particle state -- then $|\psi\rangle$ has nonzero overlap with the state $\Phi_D(0)|0\rangle$, which can be interpreted also as an approximate one particle state, where the particle is at $\vec x=0$ at time $t=0$.    Here, we also take the state $|\psi\rangle$ to be dressed, so that it is weakly annihilated by the constraints.  The overlap $\langle 0|\Phi_D(0)|\psi\rangle$ gives the amplitude for the state $|\psi\rangle$ to have a particle at the origin at time $t=0$.  This amplitude also depends on the form of the dressing for $|\psi\rangle$, and on the choice of dressing of $\Phi_D$.  A similar statement is expected to be true when $\Phi_D$ is replaced by $\Phi_D(0,R)$:  the state $\Phi_D(0,R)|0\rangle$ is a state with a charged shell at $r=R$ at time $t=0$, and this can have nonzero overlap with another state $|\psi_S\rangle$ describing a charged shell.
				
Given this, let us consider the matrix element of the commutator \eqref{NDcom} between a state $|\psi,\psi_S\rangle$ describing both an incoming particle and  shell, and the vacuum,
\beq\label{CME}
\langle 0| [\Phi_N(0),\Phi_D(0,R)] |\psi,\psi_S\rangle = - i \langle0|\phi(0)|\psi\rangle \int _N d\Sigma^\mu\, q {\check F}_{\mu\nu} A^\nu_D+ \calo(q^2)\ .
\eeq
Here
\beq
 \int_0^{-\infty} du \int r^2 d\Omega\,  u^\mu = -\int_N  d\Sigma^\mu 
 \eeq
 defines integration over the null cone, and 
\beq\label{ADdef}
  { A}_D^\nu(x)=\langle0|[A^\nu(x),\Phi_D(0,R)] |\psi_S\rangle\ .
\eeq

The Heisenberg equations for the coupled EM-matter system include the operator version of Maxwell's equation
\beq
\partial^\nu F_{\mu\nu} -\frac{1}{\alpha} \partial_\mu \partial^\nu A_\nu\equiv L_\mu{}^\nu A_\nu = j_\mu\ ,
\eeq
where $j_\mu$ is the $\phi$ current.  Acting on $ A_D$ with the differential operator $L$ thus gives
\beq\label{Ateqn}
L_{\mu\nu} { A}_D^\nu(x) = \langle0|[j_\mu(x),\Phi_D(0,R)] |\psi_S\rangle\equiv \tilde \j_\mu \ .
\eeq
This means that $ A_D$ can be interpreted as the dressing gauge potential arising from the conserved current of the evolving shell.   The source current $\tilde \j_\mu$ on the right hand side of \eqref{Ateqn} clearly will depend on the state $|\psi_S\rangle$, which contains information about the history of how the shell got to $r=R$ at $t=0$.

The integral in \eqref{CME} can be rewritten, using Gauss' theorem, as an integral over the spacelike slice at $t=0$, a surface term at $r=\infty$, and a
spacetime integral over the wedge region $W$ between the $t=0$ slice and the null cone (see fig.~\ref{fig:Wgeom}),
\beq\label{Sumint}
\int _N d\Sigma^\mu {\check F}_{\mu\nu} A^\nu_D = \int_{t=0} d^3x {\check F}_{0\nu} A^\nu_D + \int_{r=\infty} d\Sigma^\mu {\check F}_{\mu\nu} A^\nu_D - \int_W d^4 x \partial^\mu\left( {\check F}_{\mu\nu} A^\nu_D\right) \ .
\eeq
The first term vanishes, since $A^i_D(t=0)$ vanishes, from the equal-time commutators (compare \eqref{Coulpot}).  The form of $\check F$, from \eqref{Fdef}, gives for the asymptotic term
\beq
 \int_{r=\infty} d\Sigma^\mu {\check F}_{\mu\nu} A^\nu_D = \frac{1}{4\pi} \int dt\, d\Omega A^0_D(r=\infty)\ ,
\eeq
which is expected to vanish due to asymptotic vanishing of $A^0_D$.  (This can be checked in the example described below.)  In the third term of \eqref{Sumint}, the current contribution from $\partial \check F$ vanishes from vanishing of the integral measure.  These leave 
\beq\label{intenerg}
\int _N d\Sigma^\mu {\check F}_{\mu\nu} A^\nu_D = -\hf \int_W d^4 x  {\check F}_{\mu\nu} F^{\mu\nu}_D\ .
\eeq
This means that the matrix element of the leading-order commutator is the part of the action due to the interaction between the field due to the particle observed at the origin, and that of the shell.  

This interaction action depends on the specific form of $F^{\mu\nu}_D$, which will depend on how the shell evolves to $t=0$, and is determined by $|\psi_S\rangle$.  For a simple estimate of the matrix element, consider a shell that behaves classically, such as in the large mass limit, and simply stays fixed at $r=R$ for all time.  Specifically, assume that the conserved source current on the right of \eqref{Ateqn} takes the simple form
\beq\label{classj}
\tilde \j_\mu(x) = -q\, \delta^0_\mu\, \frac{\delta(r-R)}{4\pi R^2}\ .
\eeq
In this case, $F^{\mu\nu}_D$ is easily seen to be 
\bea\label{shellfield}
F^{0i}_D&=& q \frac{e^i}{4\pi r^2}\quad ,\quad r>R\cr
F^{0i}_D&=& 0 \quad ,\quad r<R\cr
F^{ij}_D&=&0\ .
\eea
The form of the resulting potential $A_D^\mu$ is somewhat more complicated, and is given in Appendix A.
Then the action \eqref{intenerg} is easily evaluated, giving 
\beq
\langle 0| [\Phi_N(0),\Phi_D(0,R)] |\psi,\psi_S\rangle = - i \langle0|\phi(0)|\psi\rangle \frac{q^2}{4\pi} \int_{W,\,>R}dt dr \frac{1}{r^2}+ \calo(q^2)\ 
\eeq
where the integral is over the  wedge $W$ in the $t,r$ plane, restricted to $r>R$.  This integral diverges, due to the integral of the interaction energy over infinite time, at $r=\infty$.  However, this divergence may again be regulated if the null dressing transitions to a spacelike dressing at some large $r=L$, as described at the end of section \ref{NullDr}. 

One can alternately consider the commutator with a null dressing for the shell, so that $D\rightarrow N$ in \eqref{NDcom}. This may be explored by similar methods, or alternately note that, for a spherical shell, the leading order dressing field will differ by a gauge transformation,
\beq
A_N^\mu = A_D^\mu + \partial^\mu \Lambda +\calo(q^2)\ .
\eeq
This may be substituted into \eqref{CME} to evaluate the corresponding matrix element of the commutator.  As a further generalization, the operator $\Phi_N(0)$ in \eqref{NDcom} may be replaced by a dressed shell $\Phi_N(0,R')$ at some other radius $R'<R$.

\subsubsection{Discussion}

The preceding expressions illustrate properties of commutators of null-dressed operators.  Eqs.~\eqref{CME},  \eqref{intenerg} in particular illustrate that the commutators, at leading order, are determined by the part of the action due to the interaction between the two operators considered. These commutators are well-defined, as long as one controls the behavior at infinity; the unregulated null dressings are problematic in this respect. Taken together with the previous result \eqref{Shellcommt}, this result suggests a general leading-order structure for commutators between dressed operators, namely of the form
\beq
[\Phi_1,\Phi_2]\sim i S_{12} \Phi_1\Phi_2
\eeq
where $S_{12}$ is an interaction action between the fields created by the operators.
				
\section{Commutators in gravity}

In this section we will provide initial exploration of commutators in the gravitational context, including in a black hole background, partially paralleling the preceding discussion of commutators in QED.  Specifically, consider the commutator between two shell operators at radii $R'$ and $R$ in a Schwarzschild background, where we imagine that $R'$ may be inside or outside the BH horizon, and $R$ lies outside, perhaps at a large distance.  The shell operators are described in analogy to those of the QED case, \eqref{shell}, \eqref{diracshell}, and comparing \eqref{Phidef}, are expected to take the $t=0$ form
\beq
\Phi(R) = \phi(R) + V^\mu(R) \partial_\mu  \phi(R) +\calo(\kappa^2)\ ,
\eeq
where $V^\mu$ is an angle-averaged dressing analogous to \eqref{Couldress} or \eqref{nullav}.  Specifically, the null case is an angular average of \eqref{nulldress}, and so is expected to have nontrivial components
\bea
\label{Gcone}
V_v(r) &=&\kappa \int\frac{d\Omega}{4\pi} \int_r^\infty dr'\left\{ h_{rv}(v,r',\theta^A) + \hf \int_{r'}^\infty d\rdp \left[ \partial_v h_{rr}(v,\rdp,\theta^A) + \frac{2M}{r^{\prime 2}} h_{rr}(v,\rdp,\theta^A)\right]\right\}\cr
V_r(r)&=&\frac{\kappa}{2}\int\frac{d\Omega}{4\pi} \int_r^\infty dr' h_{rr}(v,r',\theta^A)\ 
\eea
where the integrals over $r'$ are at constant $v$.
For an analog of the Dirac dressing \eqref{Couldress} in the non-trivial background, one can average over radial geodesics, {\it e.g.} from \eqref{static_dress}, although this leads to  more complicated expressions.

\begin{figure}
\begin{centering}
\includegraphics[scale=1.0]{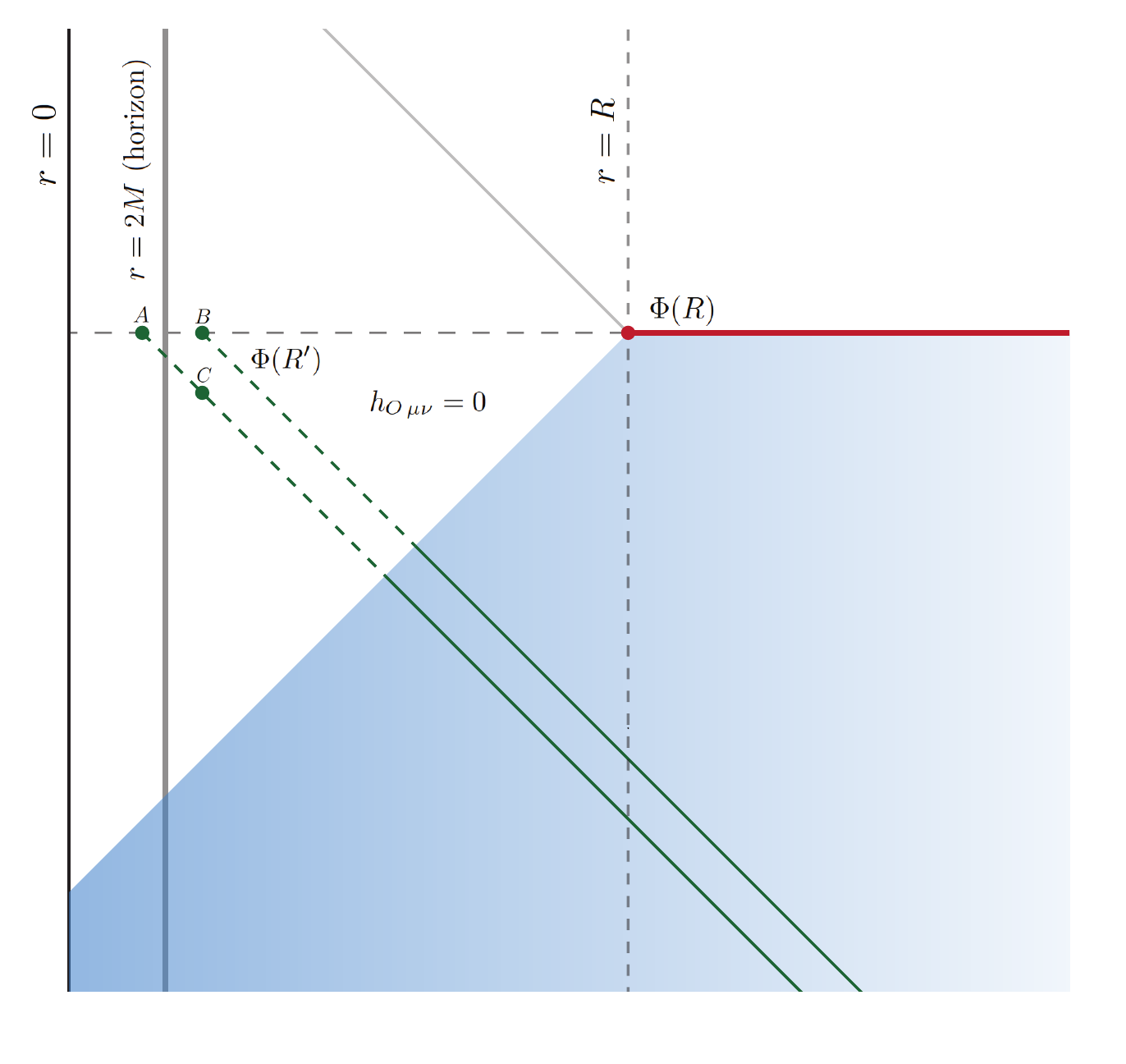}\caption{\label{fig:grav_comm}
Shown is the geometry relevant for calculating a gravitational commutator, between a null-dressed shell operator at inner radius $R'$ and another shell operator at outer radius $R$, in the background of a mass $M$ black hole geometry.  The inner operator can be placed inside the horizon, at point $A$, or outside the horizon, at points $B$ or $C$, with little effect on the commutator.  The dressing field $h_{O\mu\nu}$ of the outer operator vanishes in the left wedge, originating at  the location of the outer operator.}
\par\end{centering}
\end{figure}

\subsection{Evaluation}

In order to easily describe the transition to the interior of the horizon of the background, we assume that the innermost operator $\Phi(R')$ is dressed with such a  null-cone dressing the; outermost could be either null or spacelike dressed.  If we take the two operators to lie at equal times in a time slicing (see fig.~\ref{fig:grav_comm}), then their leading-order commutator takes the general form
\beq
[\Phi(R'),\Phi(R)] \simeq \partial_\mu\phi(R') [V^\mu(R'),\Phi(R)]\ .
\eeq

As in the QED case, we can examine the matrix element of this commutator between a state $|\psi_{R'},\psi_R\rangle$, describing the initial conditions that produce the shells, and $\langle0|$, to give
\beq\label{Gcomm}
\langle0|[\Phi(R'),\Phi(R)] |\psi_{R'},\psi_R\rangle \simeq \langle0|\partial_\mu\phi(R') |\psi_{R'}\rangle \langle0| [V^\mu(R'),\Phi(R)]|\psi_{R}\rangle\ .
\eeq
We can define the dressing field of the outer shell as
\beq
\label{grav_shell_field}
h_{O\mu\nu}= \langle0| [h_{\mu\nu},\Phi(R)]|\psi_R\rangle, 
\eeq
in which case the commutator expression in \eqref{Gcomm} becomes
\beq\label{Vfield}
\langle0| [V^\mu(R'),\Phi(R)]|\psi_{R}\rangle = V^\mu_O(R')
\eeq
and can be interpreted as the dressing of the inner shell, evaluated in the metric of the outer shell.  Specifically, it is found by substituting $h_{O\mu\nu}$ into
\eqref{Gcone}.

While explicit expressions for $h_O$ and $V_O$ are somewhat difficult to find in a black hole background, already these expressions suggest some lessons.  First, in general the first-order metric perturbation $h_O$ will vanish outside the past lightcone of the shell at $R$.  This means that the integrals \eqref{Gcone} will 
only receive contributions from the interior of this lightcone (see fig.~\ref{fig:grav_comm}).  In particular, these appear to be quite insensitive to the position of the interior shell, and specifically to whether it is inside or outside the horizon.  Thus, at least the leading order gravitational effects do not support the idea that there is an important difference between operators inside or outside a horizon, as has been previously suggested in the context of discussions of black hole complementarity\cite{STU,BankComp,BaFiComp}.  The expressions \eqref{Gcone} do not appear to locally detect the presence of the horizon.  This also strengthens the contrast with other statements, see {\it e.g.} \cite{HaOo}, regarding possible problems with defining dressed observables behind a BH horizon.

Further discussion and support for these statements could be given with more explicit expressions, most of which we defer for further work.  We do observe that
by integration by parts, the dressing \eqref{Gcone} can be also put in the form
\beq\label{Valt}
V_{Ov}(r) =  \kappa \int\frac{d\Omega}{4\pi} \int_r^\infty dr' \left[(r-r')  \left(\partial_r h_{Orv} - \hf \partial_v h_{Orr}\right) + \left(\frac{M}{r} -\frac{M}{r'}\right)h_{Orr}\right]
\eeq
up to a surface term.  
This is somewhat similar in structure to \eqref{gendress}, which did not directly generalize to a curved background, and suggests a possible generalization 
 of \eqref{gendress} that is valid in a non-flat background.

Going further seems to require determination of the properties of $h_O$.  For example, we might try to find the $h_O$ that results from an approximately classical matter shell, analogous to the classical charged shell of the preceding section.  This requires solving for propagation of gravitational perturbations in a Schwarzschild background, so is in practice challenging.

This problem can for example be solved, to explore some properties of the commutator, in a flat background, with $M=0$.  Specifically, consider a classical shell, with
stress tensor
\beq
T_{00}(x)=\frac{m}{4\pi R^{2}}\delta(r-R)\ ,
\eeq
analogous to the charged shell of \eqref{classj}.  
With initial data similar to that of \cite{DoGi1} (see eq. (64) there, and further discussion in appendix \ref{gravapp} here), we have at $t=0$
\bea
\overline{h}_{O\mu\nu} &=&\frac{\kappa m}{8\pi}\hat{r}_{\mu}\hat{r}_{\nu}\left(\frac{r-R}{r^{2}}\right)\,\Theta(r-R)\\
\partial_{t}\overline{h}_{O\mu\nu} &=& -\frac{\kappa m}{8\pi r^{2}}\left(\hat{t}_{\mu}\hat{r}_{\nu}+\hat{r}_{\mu}\hat{t}_{\nu}\right)\,\Theta(r-R)\ ,
\eea
where ${\hat r}_\mu=(0,e_i)$ and ${\hat t}_\mu=(-1,0)$.
Green function methods can be used to solve for the past and future metric perturbation (see appendix \ref{gravapp}).  
The relevant region determined by the integrals \eqref{Valt} is the region in the timelike past of the shell at $t=0$, where the metric perturbation takes the form
\bea\label{shellmet}
{\overline h}_{tt} &=& -\frac{\kappa m}{16\pi rR}(t+|r-R|)\cr
  {\overline h}_{tr} &=& -\frac{\kappa m}{32\pi r^2R}(u+R)(v+R)\cr
  \overline{h}_{rr}&=&\frac{\kappa m}{32\pi r^{3}R}\left[-2R^{2}t\log\frac{-u}{R}+r^{2}(2R+t)-2R^{2}r-t\left(3R^{2}+4Rt+t^{2}\right)\right]\cr
  {\overline h}_{ii} &=&\frac{\kappa m}{16\pi r}\left(1+\frac{R}{u}\right)  \ ;
\eea
the metric perturbation vanishes in the interior region spacelike to the shell at $t=0$ (see fig.~\ref{fig:grav_comm}).
An expression for more general regions is given in appendix \ref{gravapp}.  
With  the expression \eqref{shellmet}, and \eqref{Gcone} or \eqref{Valt} with $M=0$, one has explicit integral representations for the commutators.  These are divergent at $r=\infty$, though these divergences can be regulated by attaching a spacelike dressing as in the case of QED.  We leave further exploration and interpretation of these expressions for future work.

\subsection{Discussion: complementarity and localization}

While the preceding subsection has not explicitly evaluated  commutators in a black hole background, the leading-order structure of the commutators  for null-dressed operators is clearly revealed by the expression \eqref{Gcomm}, together with \eqref{Vfield}.

There has been considerable discussion about the role of observables inside a black hole.  An extreme viewpoint is that of ``black hole complementarity," which suggests\cite{STU,BankComp} that there is a fundamental obstruction to compatibility of observables that are inside and outside a black hole.  If this is true, that should be supported in some calculational framework, and a first place to look is at the leading-order behavior of quantized Einstein gravity.  Indeed, it was suggested by Banks and Fischler\cite{BaFiComp} that this behavior would arise in the context of solving the Wheeler-DeWitt equation.

The construction of gauge-invariant gravitational observables, such as described in this paper, directly addresses this question, since the gauge invariance can alternately be phrased in terms of these observables commuting with the constraints \eqref{constraints}, which include the Wheeler-DeWitt operator.  Thus, the leading-order form of these operators implement a leading-order solution of Wheeler-DeWitt dynamics, and  test whether there is any leading order gravitational effect of this kind.  A first observation is that while there are some subtleties and challenges with defining dressed operators in a black hole background, it appears that such gauge-invariant operators can be defined, and nothing particular dramatic happens to them at the horizon.

While the operators at $R'$ and $R$ don't commute, we find from the expressions \eqref{Gcomm} and \eqref{Vfield} that the nonzero commutator is of the same form whether or not the inner operator is inside the horizon.   For example, moving from point $A$ to point $C$ (or, $B$) of fig.~\ref{fig:grav_comm}
appears to have only minor effect on \eqref{Valt} and thus the commutator.  This suggests that if black hole complementarity is true, support for it must come from some other explanation, {\it e.g.} in nonperturbative gravity.

Of course, the statement that the operators don't commute does appear to have important consequences for the nature of locality in gravity, as has been discussed elsewhere\cite{SGalg}\cite{DoGi1,DoGi3,QFG}\cite{SBGPGS}; as with \cite{DoGi1}, the nonvanishing commutators nicely align with the statement of a ``locality bound" for gravity given in \cite{GiLia,GiLib,LQGST}.  (Similar statements have been made as a condition for a  ``code subspace\cite{ADH}" in the context of the quantum error correcting approach to holography.)  But, at least as studied with the gauge-invariant operators of this paper, these commutators appear not to significantly vary across the horizon.

A related question is that of localization of information in gravity, and whether information can in some appropriate sense be localized inside a black hole -- can a black hole be thought of as a quantum subsystem?  The existence of gravitational dressing provides an obstruction to a traditional definition of subsystems, which in quantum field theory can be phrased in terms of localized commuting observables.  Nonetheless, the fact that, at least perturbatively, information can be localized in a region such that gravitational observations outside that region are insensitive to the details of that information, suggests an alternate approach to defining subsystems in gravity\cite{QFG,DoGi4,SBGPGS}.  Then, the extension of  dressing constructions similar to those in flat space  to the black hole context suggests that the same considerations apply in this context, and that a black hole can be, at least perturbatively, thought of as properly containing information and thus behaves as a subsystem.  This is also at odds with some of the spirit of the discussion of soft hair\cite{Hawk,HPS1,HPS2,HHPS} since the
present work
 indicates that information can be localized without manifesting itself in particular in the soft part of the gravitational field\cite{DoGi4,SBGPGS}.

\section{Acknowledgements}

We thank the CERN theory group, where part of this work was carried out, for its hospitality.
This material is based in part upon work supported in part by the U.S. Department of Energy, Office of Science, under Award Number {DE-SC}0011702.  
SW was also supported by the generosity of the Len DeBenedictis Fellowship.

\appendix

\section{EM dressing field for charged shell}

In section \ref{QEDdress} we found that commutators of dressed operators depend on a dressing field, defined for example by \eqref{ADdef} in the Dirac case, and we also found that this dressing field satisfies the equation of motion \eqref{Ateqn}.  The solution $A^\mu_D$ depends on the state, that is on how the shell (or particle) evolves in time.  As an example, one can consider a shell that behaves classically, with the current $\tilde \j_\mu$ of eq. \eqref{classj} providing the right hand side of eq.~\eqref{ADdef}.  This gives a simple example illustrating aspects of the behavior of such dressing fields.

Specifically, Green function methods can be used to solve \eqref{Ateqn} given a source $\tilde \j_\mu$.  First, note that 
 as discussed in connection with \eqref{Dcomms}, $A_D$ should satisfy the Lorenz gauge condition, $\partial_\mu A_D^\mu=0$, so \eqref{Ateqn} simplifies to 
\beq
-\square A_D^\mu = \langle0|[j_\mu(x),\Phi_D(0,R)] |\psi_S\rangle = \tilde \j_\mu\ .
\eeq
Given $\tilde \j_\mu$ and initial conditions, {\it e.g.} at $t=0$, $A_D$ is determined as 
\beq\label{AGreen}
A_D^\mu(x) = \int_{t'>0} d^4 x' G(x,x') \tilde \j^\mu(x') + \int_{t'=0} d^3x' \left[-\partial_0' G(x,x') A_D^\mu(x') + G(x,x') \partial_0' A_D^\mu(x')\right]\ ,
\eeq
where $G(x,x')$ is the retarded Green function
\beq\label{Gret}
G(x,x')=\frac{\delta(t-t^{\prime}-|\mathbf{x}-\mathbf{x}^{\prime}|)}{4\pi\,|\mathbf{x}-\mathbf{x}^{\prime}|}\ ,
\eeq
which satisfies
\beq
\square^{\prime}G(x,x')=-\delta(x-x')\ .
\eeq

The needed initial conditions are found from  the equal-time commutator \eqref{ADdef} and the Lorenz gauge condition, which imply
\beq\label{Iconds}
A_D^\mu=0\quad ;\quad  E_D^i(r)=-\partial_0 A_D^i(r)=\frac{q}{4\pi}\frac{e^i}{ r^2}\Theta(r-R)\quad ,\quad \partial_0A_D^0=0\ .
\eeq

\begin{figure}
\begin{centering}
\includegraphics[scale=0.5]{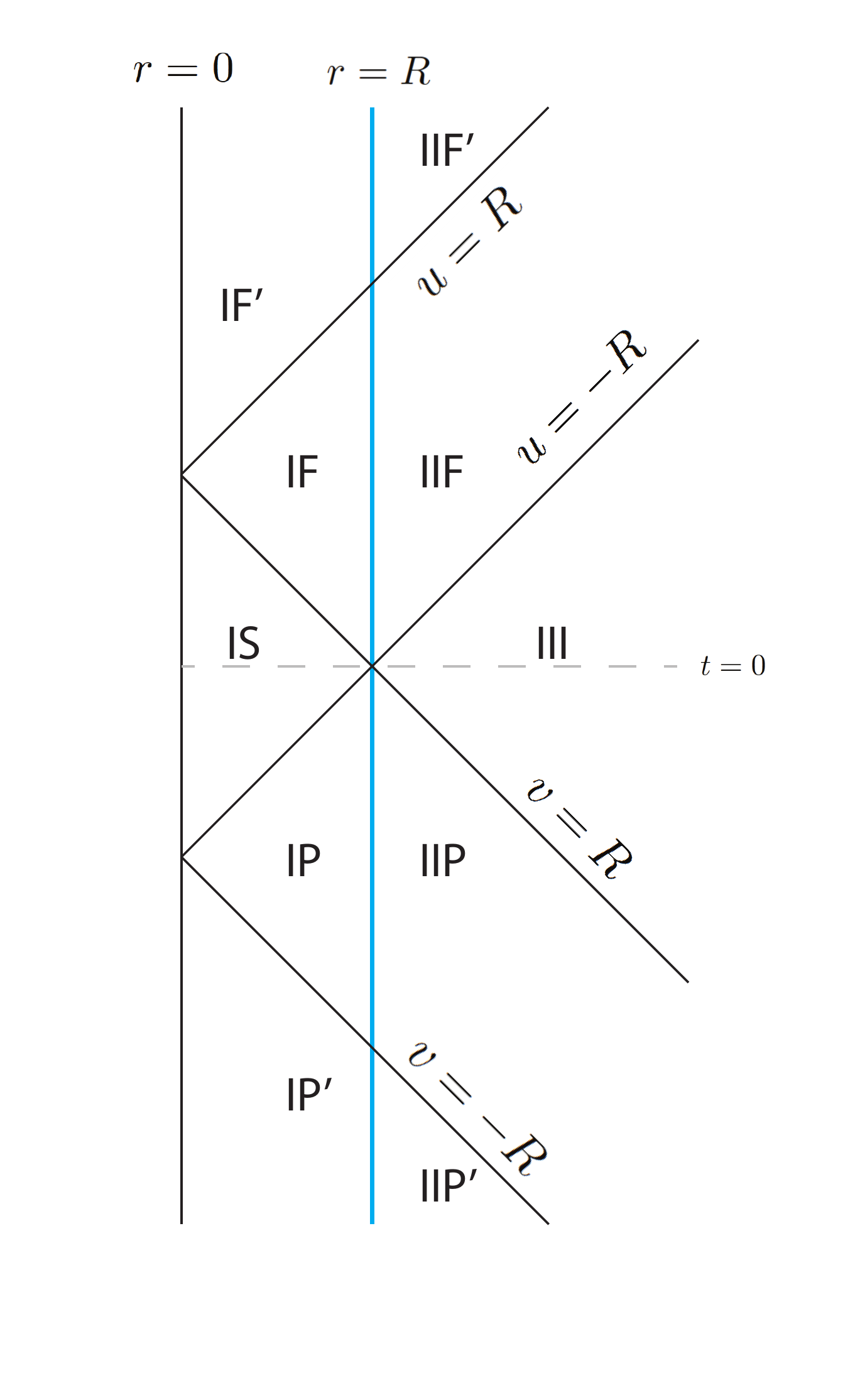}\caption{\label{fig:reg}The different regions for evaluating the gauge potential, or metric, due to a charge or matter shell at $r=R$.}
\par\end{centering}
\end{figure}

Given the classical shell $\tilde \j_\mu$ of eq.~\eqref{classj}, $A_D^\mu$  with $t>0$ may be found through explicit evaluation of \eqref{AGreen}.  The resulting $A_D$ takes different forms in the different regions shown in fig.~\ref{fig:reg}.  The corresponding results for $t<0$ are found from the 
$T$ reflection properties,
\beq
A_D^0(-t,\mathbf x) = -A_D^0(t,\vec x)\quad ,\quad A_D^i(-t,\mathbf x) = A_D^i(t,\mathbf x)\ .
\eeq

In the regions spacelike to the shell at $t=0$, only the last term of \eqref{AGreen} can contribute.  This then gives the $t>0$ results for regions IS and III
\beq
v<R\quad:\quad  A_D^\mu=0  ,
\eeq
\beq\label{AIII}
u<-R \quad:\quad A_D^0 = 0 \quad,\quad A_D^i =  -\int_{t'=0} d^3x' G(x,x') E^i_D(r') = -\frac{t q}{4\pi r^2} e^i \  .
\eeq
The latter result follows from symmetry, which ensures that only $A_D^r\neq0$, and the expression
\beq\label{Arint}
A^r_D = -\frac{q}{4\pi} \int_{t'=0} \frac{d^3x'}{4\pi} \frac{\delta(t-|\mathbf{x}-\mathbf{x}'|)}{|\mathbf{x}-\mathbf{x}'|} \frac{\ehat\cdot\ehat'}{r^{\prime 2}}\Theta(r'-R)\ .
\eeq
The integral can be evaluated by introducing the useful parameters (see \cite{SGantip})
\bea\label{params}
\mathbf{x}'' &=& \mathbf{x}'-\mathbf{x}\quad,\quad |\mathbf{x}''|=t,\cr
s &=&\frac{t}{r}\cr
\sigma &=& \left(\frac{r'}{r}\right)^{2}=1+s^{2}+2s\ehat\cdot \ehat''\ ,
\eea
and becomes
\beq
A^r_D= -\frac{q}{16\pi r} \int_{\sigma_1}^{\sigma_2} d\sigma \frac{\sigma+1-s^{2}}{2\sigma^{3/2}}\ .
\eeq
The limits depend on the region.  For $u<-R$ (region III), we have $\sigma\in\left((1-s)^{2},(1+s)^{2}\right)$, resulting in \eqref{AIII}.  The integral is also valid  in regions IF, IIF, with limits $\sigma\in\left(\frac{R^{2}}{r^{2}},(1+s)^{2}\right)$, and result
\beq
A_D^r = \frac{q}{16 \pi Rr^2}(u-R)(v-R) \ ,
\eeq
and in the far future regions IF$'$ and IIF$'$, with limits again  $\sigma\in\left((1-s)^{2},(1+s)^{2}\right)$, giving $A_D^r=0$.  

$A_D^0$ also receives a contribution from the first term of \eqref{AGreen} in regions IF, IIF, IF$'$, IIF$'$.
This is
\bea\label{A0T}
A_D^0&=& q \int_{t'>0} \frac{d^4 x'}{4\pi} \frac{\delta(t-t'-|\mathbf{x}-\mathbf{x}'|)}{|\mathbf{x}-\mathbf{x}'|}\frac{\delta(r'-R)}{4\pi R^{2}}\cr
&=& \frac{q}{8\pi} \int_{-1}^{1}d(\cos\theta')\,\frac{\Theta\left(t-\sqrt{r^{2}+R^{2}-2rR\cos\theta'}\right)}{\sqrt{r^{2}+R^{2}-2rR\cos\theta'}}\ ,
\eea
where we define $\cos\theta'=\ehat\cdot\ehat'$.   

In regions IF, IIF, with $u<R$, the $\Theta$ function cuts off the integral, and we thus find the combined result
\beq
R>u>-R\ ,\ v>R\quad:\quad A_D^0 = \frac{q}{8\pi Rr}(t-|r-R|) \quad,\quad A_D^i = \frac{q}{16\pi Rr^2}(u-R)(v-R) e^i \ .
\eeq
For $u>R$ the $\Theta$ function is irrelevant, and we find the region IF$'$, IIF$'$ result
\beq
u>R\quad :\quad A_D^0 = \frac{q}{8\pi Rr}(r+R-|r-R|)  \quad,\quad  A^i_D=0 \ .
\eeq
These expressions can be readily seen to give the field strength \eqref{shellfield}, and to obey the Lorenz gauge condition. 

\section{Gravitational dressing field for shell}\label{gravapp}

Consider perturbative gravity in flat space and fix a radius $R$.
Let $\Phi(R)$ denote a gravitationally dressed shell operator of
radius $R$, at $t=0$,
\beq
\Phi(R)=\text{\ensuremath{\phi(R)+V^{\mu}(t=0,R)\partial_{\mu}\phi(R)+{\cal O}(\kappa^{2})}}\ .
\eeq
The spacelike gravitational dressing $V$ can be found by averaging the line dressing \eqref{gline}, with radial straight line from $R$ to infinity, over all angles (see \cite{DoGi1} for a similar calculation), which gives 
\beq
V^{\mu}(t,R)=-\frac{1}{4\pi}\int_{r>R}d^{3}x\left(\frac{r-R}{r^{2}}\right)\,\gamma_{\phantom{\mu}ij}^{\mu}(t,\mathbf{x})e^i e^j\ ,
\eeq
up to a possible surface term.
Here the connection coefficients are defined by the formula \eqref{Chdef}
in terms of the metric perturbation $h$. The operator $\Phi(R)$
creates a ``dressing field'' $h_{O\mu\nu}$ defined by
\beq\label{hshelldress}
h_{O\mu\nu}(x)=\langle0|[h_{\mu\nu}(x),\Phi(R)]|\psi_{R}\rangle\ ,
\eeq
where $|\psi_{R}\rangle$ is the quantum state of the shell, 
as in equation \eqref{grav_shell_field}. 

$h_{O\mu\nu}$ can be computed
by directly solving its field equations, as was done for QED in the preceding appendix.
The trace reversed field
\beq
\overline{h}_{O\mu\nu}=h_{O\mu\nu}-\frac{1}{2}\eta_{\mu\nu}\,h_{O}\ ,
\eeq
with $h_{O}=\eta^{\mu\nu}h_{O\mu\nu}$, in the Lorenz gauge $\partial^\mu \overline h_{O\mu\nu}$,
satisfies
the equation
\beq
\square\overline{h}_{O\mu\nu}=-\frac{\kappa}{2}T_{\mu\nu}
\eeq
where $T_{\mu\nu}$ is the matrix element of the stress-energy operator
in the same states as in eq.~\eqref{hshelldress}. As a simple model, we assume that this stress tensor is that of a classical shell,
\beq
T_{00}(x)=\frac{m}{4\pi R^{2}}\delta(r-R)
\eeq
while other components vanish. The form of $\Phi$ and the commutators give initial
conditions for $\overline{h}_{O\mu\nu}$:
\beq
\overline{h}_{O\mu\nu}  =\frac{\kappa m}{8\pi}\hat{r}_{\mu}\hat{r}_{\nu}\left(\frac{r-R}{r^{2}}\right)\,\Theta(r-R)\quad,\quad
\partial_{t}\overline{h}_{O\mu\nu}  =-\frac{\kappa m}{8\pi r^{2}}\left(\hat{t}_{\mu}\hat{r}_{\nu}+\hat{r}_{\mu}\hat{t}_{\nu}\right)\,\Theta(r-R)\ .
\eeq

For $t>0$, $\overline h_O$ can be found using the retarded Green function \eqref{Gret}, in analogy with the QED case \eqref{AGreen}:
\beq\label{Ggreen}
\overline{h}_{O\mu\nu}(x) 
 = \frac{\kappa}{2}\int_{t'>0}d^{4}x'G(x,x')\,T_{\mu\nu}(x^{\prime})+\int_{t'=0}d^{3}x'\left[-\partial_{0}^{\prime}G(x,x')\overline{h}_{O\mu\nu}(x')+G(x,x')\partial_{0}^{\prime}\overline{h}_{O\mu\nu}(x')\right]\ .
\eeq
The
case of $t<0$ can be found through the T inversion symmetries
\beq
\overline{h}_{O\,tt}(-t,\mathbf{x}) =\overline{h}_{O\,tt}(t,\mathbf{x})\ ,\ 
\overline{h}_{O\,ti}(-t,\mathbf{x})  =-\overline{h}_{O\,ti}(t,\mathbf{x})\ ,\
\overline{h}_{O\,ij}(-t,\mathbf{x})  =\overline{h}_{O\,ij}(t,\mathbf{x})\ .
\eeq
We can evaluate \eqref{Ggreen} to find the various components of $\overline{h}_{O\mu\nu}$, for $t>0$.

As with the EM case, the metric perturbations given by \eqref{Ggreen} vanish in region $IS$:
\beq
v<R\quad:\quad h_{O\mu\nu}=0\ .
\eeq

$\overline{h}_{O\,tt}$ only receives a contribution from the first term in \eqref{Ggreen}, and so vanishes also in region III, with $u<-R$.  In the timelike regions to the $t=0$ shell, it is given by the integral 
\beq
\overline{h}_{Ott}(t,\mathbf{x})=\frac{\kappa m}{2}\int_{t'>0}\frac{d^{4}x'}{4\pi}\frac{\delta(t-t'-|\mathbf{x}-\mathbf{x}'|)}{|\mathbf{x}-\mathbf{x}'|}\,\frac{\delta(r'-R)}{4\pi R^{2}}\ ,
\eeq
which is the same integral as that for $A^0_D$, eq.~\eqref{A0T}.  
Similarly, we can reduce $\overline h_{Oti}$ to the EM result,
\beq
\overline{h}_{O\,tr}(t,\mathbf{x})=-\frac{\kappa m}{8\pi}\int\frac{d^{3}x'}{4\pi}\frac{1}{r'^{2}}\,\frac{\delta(t-|\mathbf{x}-\mathbf{x}'|)}{|\mathbf{x}-\mathbf{x}'|}\,\ehat \cdot\ehat' \Theta(r'-R)
\eeq
which is the same integral as \eqref{Arint} for $A_D^r$.  Collecting the results, we find for region III
\beq
u<-R\quad :\quad \overline{h}_{Ott}=0\quad , \quad \overline{h}_{O\,ti}= -\frac{\kappa m}{8\pi}\frac{t}{r^{2}} e_i\ ,
\eeq
for regions IF, IIF
\beq
R>u>-R\ ,\ v>R\quad:\quad \overline{h}_{Ott}= \frac{\kappa m}{16 \pi r R}(t-|r-R|)\quad,\quad \overline{h}_{O\,ti}=\frac{\kappa m}{32\pi r^{2}R}(u-R)(v-R)\, e_i\ ,
\eeq 
and for regions IF$'$, IIF$'$, 
\beq
u>R\quad:\quad \overline{h}_{Ott}=\frac{\kappa m}{16\pi Rr}(r+R-|r-R|)\quad ,\quad \overline{h}_{O\,ti}=0\ .
\eeq

Calculation of the spatial components $\overline h_{Oij}$ differs from that of the EM potentials.  Spherical symmetry guarantees that the spatial components take the form
\beq
\overline h_{Oij} = \alpha(r,t) e_i e_j + \beta(r,t) \delta_{ij}\ ,
\eeq
so it suffices to calculate $\overline h_{Orr}$ and the spatial trace, $\overline h_{Oii}$.

From \eqref{Ggreen}, we find 
\bea
\overline{h}_{O\,rr} & =& -\int_{t'=0}d^{3}x'\partial_{0}^{\prime}G(x,x')\,e^{i}e^{j}\overline{h}_{O\,ij}(x')\cr
 &=&\frac{\kappa m}{8\pi}\partial_{t}\left[\int_{r'>R}\frac{d^{3}x'}{4\pi}\frac{r'-R}{r'^{2}}(\ehat\cdot\ehat')^{2}\frac{\delta(t-|\mathbf{x}-\mathbf{x}'|)}{|\mathbf{x}-\mathbf{x}'|}\right].
\eea
The integral can be done like with \eqref{Arint}, by introducing the parameters $s$ and $\sigma$ of \eqref{params}, giving
\beq
\overline{h}_{O\,rr}=\frac{\kappa m}{8\pi}\partial_{t}\left[\int_{\sigma_{1}}^{\sigma_{2}}d\sigma\,\frac{(\sigma+1-s^{2})^{2}}{16\sigma}\left(\frac{1}{\sqrt{\sigma}}-\frac{R}{r\sigma}\right)\right]\ ,
\eeq
where the limits are determined by the condition $r'>R$.  For the regions III, IF$'$, or IIF$'$,  $\sigma\in\left((1-s)^{2},(1+s)^{2}\right)$, giving
\beq
u<-R\quad:\quad \overline{h}_{O\,rr}=\frac{\kappa m}{8\pi r^{2}}\left[r-R-\frac{2t^{2}}{r}+\frac{Rt}{r}\tanh^{-1}\left(\frac{t}{r}\right)\right]
\eeq
and
\beq
u>R\quad:\quad \overline{h}_{O\,rr}=\frac{\kappa m}{8\pi r^{2}}\left[-R+\frac{Rt}{r}\tanh^{-1}\left(\frac{r}{t}\right)\right]\ .
\eeq
In regions IF and IIF, $\sigma\in\left(\frac{R^{2}}{r^{2}},(1+s)^{2}\right)$ and
\beq
R>u>-R\ ,\ v>R\quad:\quad \overline{h}_{O\,rr}=\frac{\kappa m}{32\pi r^{3}R}\left[2R^{2}t\log\frac{v}{R}+r^{2}(2R-t)-2R^{2}r+t\left(3R^{2}-4Rt+t^{2}\right)\right]\ .
\eeq

The spatial trace $\overline{h}_{O\,ii}$ is likewise given
by
\beq
\overline{h}_{O\,ii}(x) = -\int_{t'=0} d^3x' \partial_0' G(x,x') \overline h_{Oii}(x')
  =\frac{\kappa m}{8\pi}\partial_{t}\left[\frac{1}{4}\int_{\sigma_{1}}^{\sigma_{2}}d\sigma\left(\frac{1}{\sqrt{\sigma}}-\frac{R}{r\sigma}\right)\right]\ .
\eeq
This yields the region III, IF, IIF, IF$'$, and IIF$'$ results
\beq
 u<-R\quad:\quad \overline{h}_{O\,ii} = \frac{\kappa m}{8\pi r}\left[1-\frac{Rr}{r^{2}-t^{2}} \right]\ ,
 \eeq
 \beq
 R>u>-R\ ,\ v>R\quad:\quad \overline{h}_{O\,ii}=\frac{\kappa m}{16\pi r}\left(1-\frac{R}{v}\right) \ ,
 \eeq
 and
 \beq
 u>R\quad ,\quad \overline{h}_{O\,ii} =- \frac{\kappa m}{8\pi r}\frac{Rr}{r^{2}-t^{2}}\ .
 \eeq

\bibliographystyle{utphys}
\bibliography{dress-gravEM}

\providecommand{\href}[2]{#2}\begingroup\raggedright\begin{thebibliography}{10}

\bibitem{UQM}
S.~B. Giddings, ``{Universal quantum mechanics},''
  \href{http://dx.doi.org/10.1103/PhysRevD.78.084004}{{\em Phys.Rev.}
  {\bfseries D78} (2008) 084004},
\href{http://arxiv.org/abs/0711.0757}{{\ttfamily arXiv:0711.0757 [quant-ph]}}.
%%CITATION = ARXIV:0711.0757;%%.

\bibitem{QFG}
S.~B. Giddings, ``{Quantum-first gravity},''
\href{http://arxiv.org/abs/1803.04973}{{\ttfamily arXiv:1803.04973 [hep-th]}}.
%%CITATION = ARXIV:1803.04973;%%.

\bibitem{QGQFA}
S.~B. Giddings, ``{Quantum gravity: a quantum-first approach},''
  \href{http://dx.doi.org/10.31526/LHEP.3.2018.01}{{\em LHEP} {\bfseries 1}
  no.~3, (2018) 1--3},
\href{http://arxiv.org/abs/1805.06900}{{\ttfamily arXiv:1805.06900 [hep-th]}}.
%%CITATION = ARXIV:1805.06900;%%.

\bibitem{Haag}
R.~Haag, {\em {Local quantum physics: Fields, particles, algebras}}.
\newblock (Texts and monographs in physics). Springer, Berlin, Germany,
1992.
\newblock
%%CITATION = INSPIRE-338216;%%.

\bibitem{Torre:1993fq}
C.~G. Torre, ``{Gravitational observables and local symmetries},''
  \href{http://dx.doi.org/10.1103/PhysRevD.48.R2373}{{\em Phys. Rev.}
  {\bfseries D48} (1993) R2373--R2376},
\href{http://arxiv.org/abs/gr-qc/9306030}{{\ttfamily arXiv:gr-qc/9306030
  [gr-qc]}}.
%%CITATION = GR-QC/9306030;%%.

\bibitem{DoGi2}
W.~Donnelly and S.~B. Giddings, ``{Observables, gravitational dressing, and
  obstructions to locality and subsystems},''
  \href{http://dx.doi.org/10.1103/PhysRevD.94.104038}{{\em Phys. Rev.}
  {\bfseries D94} no.~10, (2016) 104038},
\href{http://arxiv.org/abs/1607.01025}{{\ttfamily arXiv:1607.01025 [hep-th]}}.
%%CITATION = ARXIV:1607.01025;%%.

\bibitem{SGalg}
S.~B. Giddings, ``{Hilbert space structure in quantum gravity: an algebraic
  perspective},'' \href{http://dx.doi.org/10.1007/JHEP12(2015)099}{{\em JHEP}
  {\bfseries 12} (2015) 099},
\href{http://arxiv.org/abs/1503.08207}{{\ttfamily arXiv:1503.08207 [hep-th]}}.
%%CITATION = ARXIV:1503.08207;%%.

\bibitem{DoGi1}
W.~Donnelly and S.~B. Giddings, ``{Diffeomorphism-invariant observables and
  their nonlocal algebra},''
  \href{http://dx.doi.org/10.1103/PhysRevD.93.024030}{{\em Phys. Rev.}
  {\bfseries D93} no.~2, (2016) 024030},
  \href{http://arxiv.org/abs/1507.07921}{{\ttfamily arXiv:1507.07921
  [hep-th]}}.
[Erratum: Phys. Rev.D94,no.2,029903(2016)].
%%CITATION = ARXIV:1507.07921;%%.

\bibitem{DoGi4}
W.~Donnelly and S.~B. Giddings, ``{Gravitational splitting at first order:
  Quantum information localization in gravity},''
  \href{http://dx.doi.org/10.1103/PhysRevD.98.086006}{{\em Phys. Rev.}
  {\bfseries D98} no.~8, (2018) 086006},
\href{http://arxiv.org/abs/1805.11095}{{\ttfamily arXiv:1805.11095 [hep-th]}}.
%%CITATION = ARXIV:1805.11095;%%.

\bibitem{SBGPGS}
S.~B. Giddings, ``{Gravitational dressing, soft charges, and perturbative
  gravitational splitting},''
\href{http://arxiv.org/abs/1903.06160}{{\ttfamily arXiv:1903.06160 [hep-th]}}.
%%CITATION = ARXIV:1903.06160;%%.

\bibitem{Maroholo1}
D.~Marolf, ``{Unitarity and Holography in Gravitational Physics},''
  \href{http://dx.doi.org/10.1103/PhysRevD.79.044010}{{\em Phys. Rev.}
  {\bfseries D79} (2009) 044010},
\href{http://arxiv.org/abs/0808.2842}{{\ttfamily arXiv:0808.2842 [gr-qc]}}.
%%CITATION = ARXIV:0808.2842;%%.

\bibitem{Maroholo2}
D.~Marolf, ``{Holography without strings?},''
  \href{http://dx.doi.org/10.1088/0264-9381/31/1/015008}{{\em Class. Quant.
  Grav.} {\bfseries 31} (2014) 015008},
\href{http://arxiv.org/abs/1308.1977}{{\ttfamily arXiv:1308.1977 [hep-th]}}.
%%CITATION = ARXIV:1308.1977;%%.

\bibitem{DoGi3}
W.~Donnelly and S.~B. Giddings, ``{How is quantum information localized in
  gravity?},'' \href{http://dx.doi.org/10.1103/PhysRevD.96.086013}{{\em Phys.
  Rev.} {\bfseries D96} no.~8, (2017) 086013},
\href{http://arxiv.org/abs/1706.03104}{{\ttfamily arXiv:1706.03104 [hep-th]}}.
%%CITATION = ARXIV:1706.03104;%%.

\bibitem{GiKi}
S.~B. Giddings and A.~Kinsella, ``{Gauge-invariant observables, gravitational
  dressings, and holography in AdS},''
\href{http://arxiv.org/abs/1802.01602}{{\ttfamily arXiv:1802.01602 [hep-th]}}.
%%CITATION = ARXIV:1802.01602;%%.

\bibitem{STU}
L.~Susskind, L.~Thorlacius, and J.~Uglum, ``{The Stretched horizon and black
  hole complementarity},''
  \href{http://dx.doi.org/10.1103/PhysRevD.48.3743}{{\em Phys. Rev.} {\bfseries
  D48} (1993) 3743--3761},
\href{http://arxiv.org/abs/hep-th/9306069}{{\ttfamily arXiv:hep-th/9306069
  [hep-th]}}.
%%CITATION = HEP-TH/9306069;%%.

\bibitem{BankComp}
T.~Banks, ``{Landskepticism or why effective potentials don't count string
  models},''
\href{http://arxiv.org/abs/hep-th/0412129}{{\ttfamily arXiv:hep-th/0412129
  [hep-th]}}.
%%CITATION = HEP-TH/0412129;%%.

\bibitem{BaFiComp}
T.~Banks and W.~Fischler, ``{M theory observables for cosmological
  space-times},''
\href{http://arxiv.org/abs/hep-th/0102077}{{\ttfamily arXiv:hep-th/0102077
  [hep-th]}}.
%%CITATION = HEP-TH/0102077;%%.

\bibitem{Hawk}
S.~W. Hawking, ``{The Information Paradox for Black Holes},''
\href{http://arxiv.org/abs/1509.01147}{{\ttfamily arXiv:1509.01147 [hep-th]}}.
%%CITATION = ARXIV:1509.01147;%%.

\bibitem{HPS1}
S.~W. Hawking, M.~J. Perry, and A.~Strominger, ``{Soft Hair on Black Holes},''
  \href{http://dx.doi.org/10.1103/PhysRevLett.116.231301}{{\em Phys. Rev.
  Lett.} {\bfseries 116} no.~23, (2016) 231301},
\href{http://arxiv.org/abs/1601.00921}{{\ttfamily arXiv:1601.00921 [hep-th]}}.
%%CITATION = ARXIV:1601.00921;%%.

\bibitem{HPS2}
S.~W. Hawking, M.~J. Perry, and A.~Strominger, ``{Superrotation Charge and
  Supertranslation Hair on Black Holes},''
  \href{http://dx.doi.org/10.1007/JHEP05(2017)161}{{\em JHEP} {\bfseries 05}
  (2017) 161},
\href{http://arxiv.org/abs/1611.09175}{{\ttfamily arXiv:1611.09175 [hep-th]}}.
%%CITATION = ARXIV:1611.09175;%%.

\bibitem{HHPS}
S.~Haco, S.~W. Hawking, M.~J. Perry, and A.~Strominger, ``{Black Hole Entropy
  and Soft Hair},'' \href{http://dx.doi.org/10.1007/JHEP12(2018)098}{{\em JHEP}
  {\bfseries 12} (2018) 098},
\href{http://arxiv.org/abs/1810.01847}{{\ttfamily arXiv:1810.01847 [hep-th]}}.
%%CITATION = ARXIV:1810.01847;%%.

\bibitem{HaOo}
D.~Harlow and H.~Ooguri, ``{Symmetries in quantum field theory and quantum
  gravity},''
\href{http://arxiv.org/abs/1810.05338}{{\ttfamily arXiv:1810.05338 [hep-th]}}.
%%CITATION = ARXIV:1810.05338;%%.

\bibitem{Heem}
I.~Heemskerk, ``{Construction of Bulk Fields with Gauge Redundancy},''
  \href{http://dx.doi.org/10.1007/JHEP09(2012)106}{{\em JHEP} {\bfseries 1209}
  (2012) 106},
\href{http://arxiv.org/abs/1201.3666}{{\ttfamily arXiv:1201.3666 [hep-th]}}.
%%CITATION = ARXIV:1201.3666;%%.

\bibitem{KaLiGrav}
D.~Kabat and G.~Lifschytz, ``{Decoding the hologram: Scalar fields interacting
  with gravity},'' \href{http://dx.doi.org/10.1103/PhysRevD.89.066010}{{\em
  Phys.Rev.} {\bfseries D89} (2014) 066010},
\href{http://arxiv.org/abs/1311.3020}{{\ttfamily arXiv:1311.3020 [hep-th]}}.
%%CITATION = ARXIV:1311.3020;%%.

\bibitem{Shab}
S.~Shabanov, ``The proper field of charges and gauge invariant variables in
  electrodynamics.'' Dubna preprint JINR-E2-92-136 (unpublished).

\bibitem{PFS}
L.~V. Prokhorov, D.~V. Fursaev, and S.~V. Shabanov, ``String-like excitations
  in quantum electrodynamics,'' {\em Theoretical and Mathematical Physics}
  {\bfseries 97} no.~3, (1993) 1355--1363.

\bibitem{HaJo}
P.~E. Haagensen and K.~Johnson, ``On the wave functional for two heavy color
  sources in Yang-Mills theory,''
\href{http://arxiv.org/abs/hep-th/9702204}{{\ttfamily arXiv:hep-th/9702204
  [hep-th]}}.
%%CITATION = HEP-TH/9702204;%%.

\bibitem{SGantip}
S.~B. Giddings, ``{Generalized asymptotics for gauge fields},''
\href{http://arxiv.org/abs/1907.06644}{{\ttfamily arXiv:1907.06644 [hep-th]}}.
%%CITATION = ARXIV:1907.06644;%%.

\bibitem{MaPo}
K.~Martel and E.~Poisson, ``{Regular coordinate systems for Schwarzschild and
  other spherical space-times},''
  \href{http://dx.doi.org/10.1119/1.1336836}{{\em Am. J. Phys.} {\bfseries 69}
  (2001) 476--480},
\href{http://arxiv.org/abs/gr-qc/0001069}{{\ttfamily arXiv:gr-qc/0001069
  [gr-qc]}}.
%%CITATION = GR-QC/0001069;%%.

\bibitem{Jaff}
D.~L. Jafferis, ``{Bulk reconstruction and the Hartle-Hawking wavefunction},''
\href{http://arxiv.org/abs/1703.01519}{{\ttfamily arXiv:1703.01519 [hep-th]}}.
%%CITATION = ARXIV:1703.01519;%%.

\bibitem{Dirac1955}
P.~A. Dirac, ``{Gauge-invariant formulation of quantum electrodynamics},''
\href{http://dx.doi.org/10.1139/p55-081}{{\em Can.J.Phys.} {\bfseries 33}
  (1955) 650}.
%%CITATION = CJPHA,33,650;%%.

\bibitem{Bucholz1982}
D.~Buchholz, ``The physical state space of quantum electrodynamics,''
  \href{http://dx.doi.org/10.1007/BF02029133}{{\em Communications in
  Mathematical Physics} {\bfseries 85} no.~1, (1982) 49--71}.

\bibitem{Steinmann1983}
O.~Steinmann, ``{Perturbative {QED} in Terms of Gauge Invariant Fields},''
\href{http://dx.doi.org/10.1016/0003-4916(84)90053-8}{{\em Annals Phys.}
  {\bfseries 157} (1984) 232}.
%%CITATION = APNYA,157,232;%%.

\bibitem{Steinmann2004}
O.~Steinmann, ``{Physical fields in QED},'' {\em Prog.Math.} {\bfseries 251}
  (2007) 301--310,
\href{http://arxiv.org/abs/hep-th/0411095}{{\ttfamily arXiv:hep-th/0411095
  [hep-th]}}.
%%CITATION = HEP-TH/0411095;%%.

\bibitem{Mand}
S.~Mandelstam, ``{Quantum electrodynamics without potentials},''
\href{http://dx.doi.org/10.1016/0003-4916(62)90232-4}{{\em Annals Phys.}
  {\bfseries 19} (1962) 1--24}.
%%CITATION = APNYA,19,1;%%.

\bibitem{CPA}
S.~Choi, S.~S. Pradhan, and R.~Akhoury, ``{Supertranslation Hair of
  Schwarzschild Black Hole: A Wilson Line Perspective},''
\href{http://arxiv.org/abs/1910.05882}{{\ttfamily arXiv:1910.05882 [hep-th]}}.
%%CITATION = ARXIV:1910.05882;%%.

\bibitem{GiLia}
S.~B. Giddings and M.~Lippert, ``{Precursors, black holes, and a locality
  bound},'' \href{http://dx.doi.org/10.1103/PhysRevD.65.024006}{{\em Phys.Rev.}
  {\bfseries D65} (2002) 024006},
\href{http://arxiv.org/abs/hep-th/0103231}{{\ttfamily arXiv:hep-th/0103231
  [hep-th]}}.
%%CITATION = HEP-TH/0103231;%%.

\bibitem{GiLib}
S.~B. Giddings and M.~Lippert, ``{The Information paradox and the locality
  bound},'' \href{http://dx.doi.org/10.1103/PhysRevD.69.124019}{{\em Phys.Rev.}
  {\bfseries D69} (2004) 124019},
\href{http://arxiv.org/abs/hep-th/0402073}{{\ttfamily arXiv:hep-th/0402073
  [hep-th]}}.
%%CITATION = HEP-TH/0402073;%%.

\bibitem{LQGST}
S.~B. Giddings, ``{Locality in quantum gravity and string theory},''
  \href{http://dx.doi.org/10.1103/PhysRevD.74.106006}{{\em Phys.Rev.}
  {\bfseries D74} (2006) 106006},
\href{http://arxiv.org/abs/hep-th/0604072}{{\ttfamily arXiv:hep-th/0604072
  [hep-th]}}.
%%CITATION = HEP-TH/0604072;%%.

\bibitem{ADH}
A.~Almheiri, X.~Dong, and D.~Harlow, ``{Bulk Locality and Quantum Error
  Correction in AdS/CFT},''
  \href{http://dx.doi.org/10.1007/JHEP04(2015)163}{{\em JHEP} {\bfseries 04}
  (2015) 163},
\href{http://arxiv.org/abs/1411.7041}{{\ttfamily arXiv:1411.7041 [hep-th]}}.
%%CITATION = ARXIV:1411.7041;%%.

\end{thebibliography}\endgroup

\end{document}